\documentclass[twocolumn]{aa}
\usepackage{graphicx}
\usepackage{txfonts}

\def\ms{\,m\,s$^{-1}$}         
\def\kms{\,km\,s$^{-1}$}       
\def\vsini{$v$\,sin\,$i$}      
\def\Msol{M_\odot}             

\begin{document}
   \title{Doppler follow-up of OGLE transiting companions \\
   in the Galactic bulge\thanks{Based on observations 
	   collected with the UVES and FLAMES spectrographs at the VLT/UT2 Kueyen telescope 
	   (Paranal Observatory, ESO, Chile: program 70.C-0209 and 71.C-0251)}
	   }

   \author{F. Bouchy\inst{1,4}, F. Pont\inst{1,4}, C. Melo\inst{2}, N.C. Santos\inst{3,4},
           M. Mayor\inst{4}, D. Queloz\inst{4} \and S. Udry\inst{4}
          }

   \offprints{Francois.Bouchy@oamp.fr}

   \institute{Laboratoire d'Astrophysique de Marseille, 
               Traverse du Siphon, BP8, 13376 Marseille Cedex 12, France
         \and
	      European Southern Observatory, Casilla 19001, Santiago 19, Chile  
         \and
	     Centro de Astronomia e Astrof\'{\i}sica da Universidade de Lisboa, Tapada da Ajuda, 1349-018 Lisboa, Portugal     
	  \and
              Observatoire de Gen\`eve, 51 ch. des Maillettes, 1290 Sauverny, Switzerland
            }

   \date{Received ; accepted }

   \abstract{Two years ago, the OGLE-III survey (Optical Gravitational Lensing 
Experiment) announced the detection of 54 short period multi-transiting objects 
in the Galactic bulge (Udalski et al., \cite{udalski1}, \cite{udalski2}). Some of these 
objects were considered to be potential hot Jupiters. In order to determine the true 
nature of these objects and to characterize their actual mass, we conducted a radial 
velocity follow-up of 18 of the smallest transiting candidates. We describe here our 
procedure and report the characterization of 8~low mass star transiting companions, 
2~grazing eclipsing binaries, 2~triple systems, 1~confirmed exoplanet (OGLE-TR-56b), 
1~possible exoplanet (OGLE-TR-10b), 1~clear false positive and 3~unsolved cases. 
The variety of cases encountered in our follow-up covers a large part of the 
possible scenarii occuring in the search for planetary transits. As a by-product, 
our program yields precise masses and radii of low mass stars.

   }

   \authorrunning{F. Bouchy et al.}
   \titlerunning{Doppler follow-up of OGLE}

   \maketitle

\section{Introduction}

Since 1995 the search for planets by radial velocity surveys has lead to the 
detection of more than 120 planetary candidates. The diversity of orbital 
characteristics, the mass distribution of the planets (actually only the $m\sin i$) 
and its link with brown dwarfs and low-mass stars as well as the characteristics 
of host stars prompted a reexamination of planetary formation theory (e.g., Udry et al. 
\cite{udry}, Santos et al. \cite{santos03}, Eggenberger et al. \cite{eggen}). 
The most unexpected fact was the existence of extrasolar giant planets (EGPs) in very short orbits. 
Extra mechanisms, not envisioned by the study of our Solar system, have been 
suggested to explain these objects, like the migration of planets in the proto-planetary 
disk and gravitational interactions (e.g., Goldreich \& Tremaine \cite{goldreich}, Lin et al. 
\cite{lin}). 

Monitoring of photometric transits caused by an EGP passing on the disk of its 
hosting star and obscuring part of its surface provides an opportunity to determine 
its actual size. When combined with spectroscopic observations, it leads to the unambiguous 
characterization of the two fundamental parameters (mass and radius) used for 
internal structure studies of EGPs.
The discovery of HD209458 by both Doppler measurements (Mazeh et al. \cite{mazeh}) and 
photometric transit (Charbonneau et al. \cite{charbonneau}; Henry et al. \cite{henry}) 
led to the first complete characterization of an EGP, 
illustrating the real complementarity of the two methods. 
These last years many extensive ground-based photometric programs have been initiated 
to detect transits by short period EGPs (Horne \cite{horne}). The OGLE-III survey 
(Optical Gravitational Lensing Experiment) announced recently the detection of 137 
short-period multi-transiting objects (Udalski et al., \cite{udalski1}, \cite{udalski2}, 
\cite{udalski3}, \cite{udalski4}). The estimated radius of these 
objects range from 0.5 Jupiter radius to 0.5 solar radius and their orbital periods range 
from 0.8 to 8 days. The smallest objects could be suspected to be EGPs, but 
considering only the radius measured by OGLE, one can not conclude on the planetary 
nature of the objects per se. They could as well be brown-dwarf or low-mass stars since 
in the low mass regime the radius is independent of the mass (Guillot, \cite{guillot}). 
No information on the mass of these companions is given by the transit measurements.
Doppler follow-up of these candidates is the only way to confirm the 
planetary, brown-dwarf or low-mass star nature of the companions. 
Planetary transit detection suffers also some ambiguity related to the 
configuration of the system. The radial velocity measurement is therefore 
very important in that respect to discriminate true central transits from other cases 
due to, for example, grazing eclipsing binaries, blended system and stellar activity.
The spectroscopy of the central star, which is a by-product of radial velocity 
measurement, is necessary to constrain the star radius and therefore the companion 
radius. The measurement of the true mass of a companion by the radial velocity orbit, 
coupled with the measurement of its radius, lead to a direct measurement of its mean 
density, an essential parameter for the study of internal structure of EGPs, 
brown dwarfs and low mass stars. 

The difficulties of Doppler follow-up of OGLE candidates come from the faintness of the 
stars (with V magnitudes in the range 14-18). The fields, which are located in the 
galactic disk, are furthermore very crowed. To characterize a hot Jupiter, one needs 
radial velocity precision better than 100 {\ms} and the capability to distinguish 
whether the system is blended or not by a third star like the triple system HD41004 
(Santos et al., \cite{santos03}).

Several teams are involved in the Doppler follow-up of OGLE candidates. 
Konacki et al. (\cite{konacki1}) announced 
first that the companion of OGLE-TR-56 is a planet of 0.9 Jupiter mass. Dreizler et al. 
(\cite{dreizler2}) gave an upper limit of 2.5 Jupiter mass for the companion of 
OGLE-TR-3. This object was however refuted by Konacki et al. (\cite{konacki2}) who 
also gave informations on 3 other OGLE companions (OGLE-TR-10, 33 and 58). Additional 
measurements, conducted by Torres et al. (\cite{torres}), lead to improve the mass 
determination of OGLE-TR-56b to 1.45 Jupiter mass. Recently, we announced the 
characterization of planets OGLE-TR-113b, OGLE-TR-132b (Bouchy et al. \cite{bouchy}, 
Moutou et al. \cite{moutou}) and OGLE-TR-111b (Pont et al. \cite{pont}).

We present in this paper the Doppler follow-up observations of 18 OGLE multi-transiting 
companions (OGLE-TR-5, 6, 7, 8, 10, 12, 17, 18, 19, 33, 34, 35, 48, 49, 55, 56, 
58 and 59) from the 54 detected in the Galactic bulge (Udalski et al., \cite{udalski1}, 
\cite{udalski2}).

\section{Target selection and observations}

\subsection{UVES}
\begin{figure}
\resizebox{8.5cm}{!}{\includegraphics{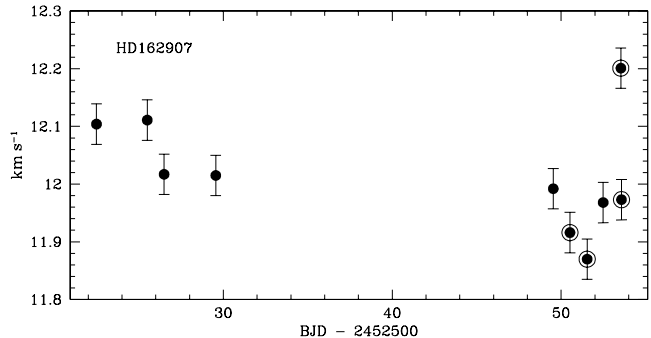}}
\caption{Doppler measurements of the standard star HD162907 made with UVES. The 
dispersion of 93 {\ms} is dominated by the centering error. Encircled points correspond 
to measurements made with a seeing lower than 0.9 arcsec.} 
\label{hd162907}
\end{figure}

We have obtained 16 hours in service mode with the UVES spectrograph on the ESO-VLT in 
October 2002 (program 70C.0209A). For this run we selected 
3 candidates (OGLE-TR-8, 10 and 12) with an estimated companion radius less than 1.6 
Jupiter radius (following the value given by Udalski et al., \cite{udalski1}). We were 
aware that UVES has no fiber to produce a stable illumination at the entrance of the 
spectrograph and that the Iodine cell method is unusable for such faint stars. However, 
using the smallest slit (0.3 arcsec) in a medium seeing condition (0.9-1.4 arcsec), 
we can minimize the velocity error stemming from the shift of the photo-center on the slit. 
Considering as well the very good guiding ($<0.1$ arcsec) and the centering accuracy of UT2 
(about $0.1$ arcsec), we simulated that for a 1.0 arcsec seeing one should reach an overall 
stabilization of the photo-center of about 1/70 of the slit width. This corresponds to a 
radial velocity error of about 40 {\ms}. In order to check our accuracy we added the 
bright standard radial velocity star HD162907 (selected from the CORALIE exoplanet survey) 
close to our 3 candidates. Moreover, in order to track and to correct 
instrumental calibration drifts we took Thorium exposures before and after each science 
exposures. With such a procedure we expected an overall radial velocity precision of 
about 50 {\ms}. We used the red arm of the spectrograph with a central wavelength of 
580 nm. Notice that in order to reach this precision it is mandatory not to 
conduct this program in good seeing conditions (i.e seeing less than 0.9 arcsec). 
We noticed that such an unusual requirement is extremely rare in Paranal. 
We made 8 measurements of each targets with exposure time between 20 and 40 minutes.
Figure~\ref{hd162907} shows the result obtained with UVES on HD162907. The dispersion 
of 93 {\ms} is dominated by the centering error especially during night with good seeing 
condition. If we eliminate the 4 measurements made with a seeing lower than 0.9 arcsec, 
the dispersion reaches 54 {\ms}. 
    
\subsection{FLAMES}
\begin{figure*}
\resizebox{16cm}{!}{\includegraphics{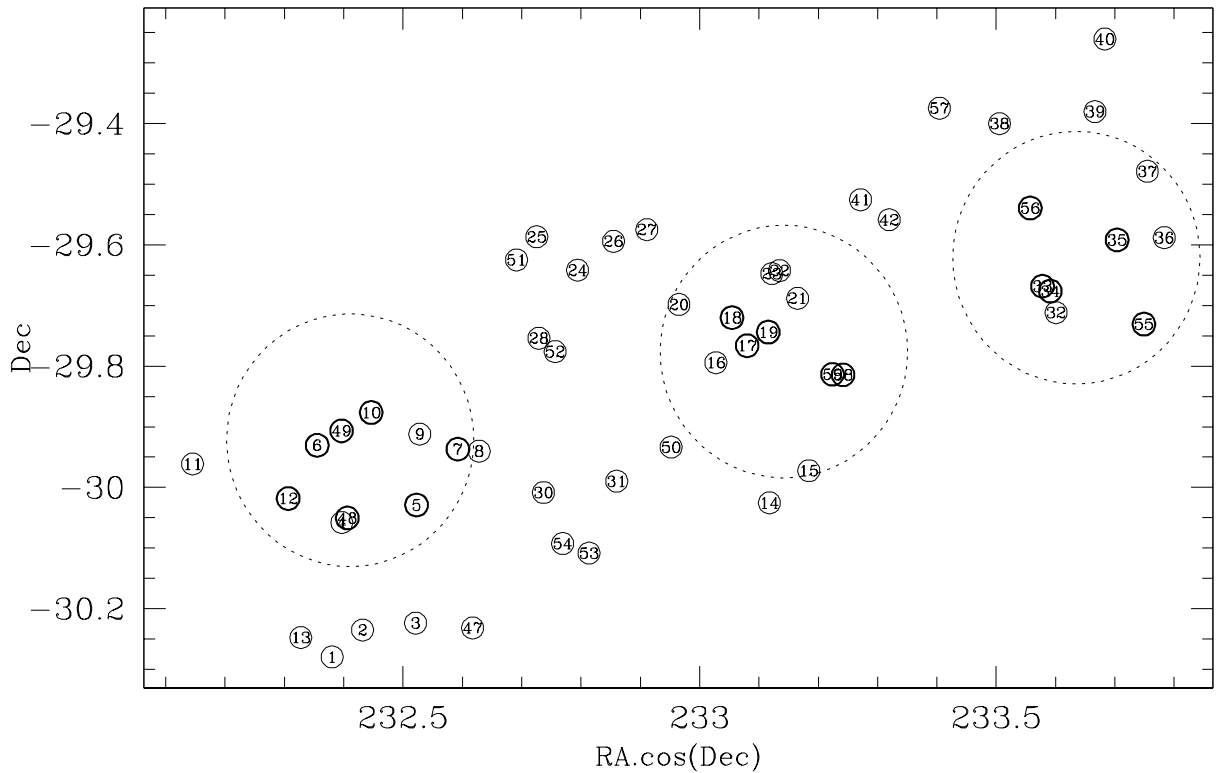}}
\caption{Positions of the 54 OGLE candidates on the sky and localization of our 3 
selected FLAMES fields. Bold circles correspond to the 17 observed OGLE candidates 
during our run.} 
\label{fields}
\end{figure*}

The FLAMES facilities on the VLT (available since march 2003) seems to be 
the most efficient way to conduct the Doppler follow-up of OGLE candidates. 
FLAMES is a multi-fiber link which allows to feed the spectrograph UVES with up to 
7 targets on a field-of-view of 25 arcmin diameter in addition with the 
simultaneous thorium calibration. The fiber link allows for a stable 
illumination at the entrance of the spectrograph and the simultaneous 
thorium calibration is used to track instrumental drift. Forty five minutes on a 17 
magnitude star reach a signal-to-noise ratio of about 8, corresponding to a photon noise 
uncertainty of about 25 {\ms} on a non-rotating K dwarf star.
We have obtained 24 hours in service mode on this instrument (program 71.C-0251A) in 
order to observe in May and June 2003 17 candidates (OGLE-TR-5, 6, 7, 10, 12, 17, 18, 
19, 33, 34, 35, 48, 49, 55, 56, 58 and 59) located in 3 FLAMES fields 
(see Fig.~\ref{fields}). For this program, we selected a sample including 7 candidates 
with an estimated companion radius less than 1.6 Jupiter radius (following the value 
given by Udalski et al., \cite{udalski1}, \cite{udalski2}) and completed each fields 
with targets having larger companion radius in order to bring constraint 
in the mass-radius relation of low-mass stars. 

\subsection{HARPS}
HARPS is a new fiber-fed spectrograph on the ESO 3.6-m telescope (Mayor et al. \cite{mayor}) 
dedicated to high-precision Doppler measurements. 
During the second commissioning in June 2003, we tested the capability of the instrument 
to realize radial velocity measurements on faint stars and made 5 measurements of 1-hour 
exposure on OGLE-TR-56. The 20 first blue spectral orders were not used to compute 
radial velocity due to their too low S/N ($<1$).

\section{Spectroscopic analysis}

\subsection{Radial velocities}

The spectra obtained from the FLAMES and UVES spectrographs were extracted using 
the standard ESO-pipeline with bias, flat-field and background correction. Wavelength 
calibration was performed with ThAr spectra. The radial velocities were obtained by 
weighted cross-correlation with a numerical mask constructed from the Sun spectrum atlas. 
In the case of FLAMES and HARPS, the simultaneous ThAr spectrum was used in order to 
compute the instrumental drift by cross-correlation with a Thorium mask. Radial velocity 
uncertainties (in {\kms}) were computed as a function of the SNR per pixel of the spectrum, the 
width (FWHM in {\kms}) and depth (C in \%) of the Cross-Correlation Function (CCF) through the 
following relation based on photon noise simulations:

\[
\sigma_{\rm RV} = \frac{3 \cdot \sqrt{\rm FWHM}}{\rm SNR \cdot C}\,.
\]

However, our measurements are clearly not photon noise limited and we added 
quadratically an empirically determined uncertainty of 35 {\ms} in order to take into the account systematic 
errors probably due to wavelength calibration errors, fiber-to-fiber contamination, and 
residual cosmic rays. We checked that, on non-rotating dwarf stars, the O-C residuals of 
our measurements are in agreement with the estimated uncertainties based on this 
relation. Our best O-C residuals indicate that we reached radial velocity precision 
lower than 100 {\ms}.\\

Our radial velocity measurements and Cross-Correlation Function parameters are listed 
in Table~\ref{tablevr}. For some candidates observed with FLAMES at very low SNR (SNR$< 5$), 
the depth of the CCF is correlated with the SNR. This is a clear indication that 
the spectra are contaminated by background light due in part to the fiber-to-fiber 
contamination. We take into consideration this effect in order to correctly estimate 
the SNR and to compute our radial velocity uncertainties more strictly.    
Phase-folded radial velocities and results are presented and 
discussed in section~\ref{results}. 

\begin{table}
\caption{Radial velocity measurements (in the barycentric frame) and CCF parameters. Label $a$ and $b$ 
indicate that 2 components are present in the CCF. BJD in the range [522-557], [750-798] and 
[809-812] correspond respectively to UVES, FLAMES and HARPS measurements.}
\label{tablevr}
\begin{tabular}{r r r r r r} \hline \hline
BJD & RV & depth & FWHM & SNR & $\sigma_{\rm RV}$  \\ 
{[$-2452000\,$d]} & [{\kms}] & [\%] & [{\kms}] & & [{\kms}] \\ \hline 
OGLE-5 & &  & & & \\ \hline
759.72111  &  $-$15.452  &  1.88  &  119.4 &  8.9 & 1.959 \\ 
764.85708  &  13.273  &  2.17  &  151.5 &  7.6 & 2.239 \\ 
766.79458  &  48.733  &  2.28  &  116.7 &  11.6 & 1.226 \\ 
767.72839  &  12.256  &  1.83  &  142.5 &  8.8 & 2.224 \\ 
768.84710  &  $-$11.662  &  2.56  &  119.0 &  15.5 & 0.825 \\ 
769.85598  &  55.993  &  2.33  &  121.4 &  10.0 & 1.419 \\ 
770.82050  &  60.894  &  2.34  &  122.9 &  11.0 & 1.293 \\ 
797.76880  &  $-$34.035  &  2.40  &  123.3 &  11.3 & 1.229 \\ \hline
OGLE-6 & &  & & & \\ \hline
759.72111  &  33.160  &  7.34  &  35.4 &  5.0 & 0.488 \\ 
764.85708  &  46.860  &  9.95  &  34.9 &  6.4 & 0.281 \\ 
766.79458  &  $-$10.166  &  10.93  &  34.0 &  7.1 & 0.228 \\ 
767.72839  &  $-$10.385  &  8.99  &  33.2 &  5.6 & 0.345 \\ 
768.84710  &  33.995  &  13.55  &  38.0 &  10.5 & 0.135 \\ 
769.85598  &  44.341  &  10.22  &  36.2 &  6.1 & 0.292 \\ 
770.82050  &  8.344  &  12.33  &  35.0 &  7.2 & 0.203 \\ 
797.76880  &  24.197  &  11.86  &  36.6 &  7.5 & 0.207 \\ \hline
OGLE-7 & &  & & & \\ \hline
759.72111  &  8.454  &  6.94  &  55.3 &  9.6 & 0.337 \\ 
764.85708  &  20.473  &  7.13  &  54.6 &  10.4 & 0.301 \\ 
766.79458  &  $-$11.924  &  8.32  &  54.8 &  11.3 & 0.239 \\ 
767.72839  &  14.997  &  6.79  &  48.6 &  9.5 & 0.326 \\ 
768.84710  &  $-$42.394  &  8.04  &  54.7 &  14.4 & 0.195 \\ 
769.85598  &  10.541  &  7.13  &  50.4 &  10.9 & 0.276 \\ 
770.82050  &  $-$6.034  &  7.48  &  50.6 &  11.2 & 0.257 \\ 
797.76880  &  9.175  &  7.30  &  52.4 &  12.0 & 0.250 \\  \hline
OGLE-8a & &  & & & \\ \hline
522.55485  &  19.193  &  10.94  &  16.9 &  6.9 & 0.167 \\ 
526.55409  &  69.882  &  11.60  &  18.1 &  8.6 & 0.133 \\ 
549.59728  &  29.917  &  10.50  &  16.8 &  4.1 & 0.288 \\ 
550.58081  &  $-$54.831  &  11.04  &  17.3 &  5.1 & 0.224 \\ 
551.59645  &  $-$78.950  &  11.02  &  17.7 &  5.9 & 0.197 \\ 
552.54163  &  $-$17.312  &  11.72  &  17.9 &  7.2 & 0.154 \\ 
553.59422  &  64.676  &  10.67  &  17.4 &  5.9 & 0.202 \\ 
556.50991  &  $-$79.423  &  10.34  &  17.0 &  4.2 & 0.287 \\ \hline
\end{tabular}
\end{table}

\begin{table}
\begin{tabular}{r r r r r r}\hline \hline
BJD & RV & depth & FWHM & SNR & $\sigma_{\rm RV}$  \\ 
{[$-2452000\,$d]} & [{\kms}] & [\%] & [{\kms}] & & [{\kms}] \\ \hline 
OGLE-8b & &  & & & \\ \hline
522.55485  &  $-$27.289  &  10.33  &  16.1 &  6.9 & 0.172 \\ 
526.55409  &  $-$80.238  &  10.35  &  16.7 &  8.6 & 0.142 \\ 
549.59728  &  $-$37.875  &  9.54  &  17.3 &  4.1 & 0.321 \\ 
550.58081  &  48.938  &  9.70  &  17.1 &  5.1 & 0.253 \\ 
551.59645  &  74.409  &  10.02  &  17.7 &  5.9 & 0.216 \\ 
552.54163  &  10.650  &  9.87  &  17.2 &  7.2 & 0.179 \\ 
553.59422  &  $-$74.939  &  10.29  &  15.4 &  5.9 & 0.197 \\ 
556.50991  &  74.659  &  9.14  &  15.4 &  4.2 & 0.309 \\ \hline
OGLE-10 & &  & & & \\ \hline
522.50823  &  $-$6.421  &  30.62  &  11.7 &  6.1 & 0.065 \\ 
525.50905  &  $-$6.413  &  30.14  &  11.9 &  6.8 & 0.061 \\ 
526.51146  &  $-$6.485  &  29.06  &  10.7 &  6.9 & 0.060 \\ 
549.55329  &  $-$6.607  &  29.94  &  11.9 &  4.4 & 0.086 \\ 
550.53700  &  $-$6.135  &  29.86  &  11.5 &  3.4 & 0.106 \\ 
551.55409  &  $-$6.475  &  30.28  &  11.2 &  3.8 & 0.094 \\ 
552.50287  &  $-$6.410  &  31.03  &  11.9 &  6.4 & 0.063 \\ 
553.54729  &  $-$6.135  &  30.58  &  11.3 &  6.4 & 0.062 \\ 
759.72111  &  $-$6.291  &  24.03  &  12.9 &  8.4 & 0.064 \\ 
764.85708  &  $-$6.236  &  27.18  &  13.1 &  10.9 & 0.051 \\ 
766.79458  &  $-$6.212  &  26.17  &  13.1 &  9.8 & 0.055 \\ 
767.72839  &  $-$6.274  &  25.11  &  12.5 &  9.0 & 0.059 \\ 
768.84710  &  $-$6.365  &  27.19  &  12.8 &  14.6 & 0.044 \\ 
769.85598  &  $-$6.238  &  25.91  &  12.9 &  9.9 & 0.055 \\ 
770.82050  &  $-$6.105  &  26.89  &  12.5 &  15.3 & 0.043 \\ 
797.76880  &  $-$6.130  &  26.80  &  12.6 &  11.2 & 0.050 \\ \hline
OGLE-12 & &  & & & \\ \hline
522.52429  &  $-$10.897  &  27.06  &  11.8 &  8.0 & 0.059 \\ 
525.52490  &  43.661  &  27.50  &  11.7 &  8.1 & 0.058 \\ 
526.52813  &  57.877  &  27.47  &  12.0 &  9.0 & 0.055 \\ 
549.57081  &  $-$0.796  &  26.78  &  12.3 &  5.4 & 0.081 \\ 
550.55534  &  21.265  &  27.57  &  11.5 &  4.3 & 0.093 \\ 
551.57020  &  45.212  &  27.21  &  11.6 &  7.0 & 0.064 \\ 
552.51801  &  58.079  &  27.61  &  12.2 &  7.4 & 0.062 \\ 
553.56592  &  54.157  &  26.78  &  12.3 &  5.6 & 0.078 \\ 
759.72111  &  52.149  &  20.08  &  12.9 &  6.6 & 0.089 \\ 
764.85708  &  $-$10.531  &  22.86  &  12.4 &  8.0 & 0.068 \\ 
766.79458  &  16.577  &  23.38  &  13.3 &  9.3 & 0.061 \\ 
767.72839  &  39.484  &  21.31  &  12.8 &  8.2 & 0.071 \\ 
768.84710  &  57.534  &  24.75  &  13.2 &  15.4 & 0.045 \\ 
769.85598  &  56.098  &  22.96  &  12.6 &  8.0 & 0.068 \\ 
770.82050  &  39.236  &  24.05  &  13.2 &  11.6 & 0.052 \\ 
797.76880  &  15.003  &  23.58  &  12.8 &  9.4 & 0.060 \\ \hline
\end{tabular}
\end{table}

\begin{table}
\begin{tabular}{r r r r r r}\hline \hline
BJD & RV & depth & FWHM & SNR & $\sigma_{\rm RV}$  \\ 
{[$-2452000\,$d]} & [{\kms}] & [\%] & [{\kms}] & & [{\kms}] \\ \hline 
OGLE-17 & &  & & & \\ \hline
759.75994  &  $-$64.941  &  26.73  &  10.0 &  7.7 & 0.058 \\ 
764.81842  &  $-$9.416  &  26.36  &  10.3 &  6.1 & 0.069 \\ 
766.84092  &  $-$2.377  &  26.49  &  9.8 &  6.9 & 0.062 \\ 
767.76871  &  $-$8.671  &  17.55  &  9.3 &  3.8 & 0.142 \\ 
768.76694  &  $-$19.328  &  28.28  &  10.0 &  8.5 & 0.053 \\ 
769.81610  &  $-$32.710  &  26.32  &  10.1 &  7.5 & 0.060 \\ 
770.90035  &  $-$46.353  &  24.91  &  9.7 &  6.5 & 0.067 \\ 
791.72546  &  $-$21.462  &  26.47  &  9.9 &  7.5 & 0.059 \\ \hline
OGLE-18 & &  & & & \\ \hline
759.75994  &  $-$24.248  &  4.67  &  87.5 &  8.4 & 0.716 \\ 
764.81842  &  $-$12.247  &  4.13  &  72.2 &  7.1 & 0.870 \\ 
766.84092  &  0.683  &  4.26  &  66.8 &  7.3 & 0.789 \\ 
767.76871  &  $-$91.001  &  3.75  &  60.9 &  5.0 & 1.249 \\ 
768.76694  &  $-$8.342  &  4.55  &  78.8 &  8.9 & 0.659 \\ 
769.81610  &  $-$73.094  &  4.90  &  69.6 &  8.2 & 0.624 \\ 
770.90035  &  $-$23.657  &  4.29  &  82.5 &  7.1 & 0.895 \\ 
791.72546  &  $-$29.868  &  4.23  &  81.9 &  8.4 & 0.765 \\ \hline
OGLE-19 & &  & & & \\ \hline
759.75994  &  $-$33.380  &  25.46  &  10.0 &  5.3 & 0.078 \\ 
764.81842  &  $-$33.245  &  24.85  &  9.5 &  4.9 & 0.084 \\ 
766.84092  &  $-$33.287  &  25.14  &  9.9 &  4.8 & 0.086 \\ 
767.76871  &  $-$33.385  &  17.80  &  9.3 &  3.6 & 0.148 \\ 
768.76694  &  $-$33.087  &  29.12  &  9.9 &  5.9 & 0.065 \\ 
769.81610  &  $-$33.244  &  25.75  &  9.8 &  5.2 & 0.078 \\ 
770.90035  &  $-$33.394  &  23.78  &  10.2 &  4.6 & 0.095 \\ 
791.72546  &  $-$33.414  &  24.53  &  9.6 &  5.2 & 0.080 \\ \hline
OGLE-33a & &  & & & \\ \hline
764.89532  &  $-$28.500  &  4.00  &  61.2 &  17.2 & 0.343 \\ 
766.87982  &  $-$28.500  &  4.00  &  61.2 &  16.0 & 0.368 \\ 
767.80618  &  $-$28.500  &  4.00  &  61.2 &  15.6 & 0.378 \\ 
769.77114  &  $-$28.500  &  4.00  &  61.2 &  16.9 & 0.349 \\ 
770.85930  &  $-$28.500  &  4.00  &  61.2 &  16.1 & 0.366 \\ 
790.70926  &  $-$28.500  &  4.00  &  61.2 &  16.8 & 0.351 \\ 
797.73001  &  $-$28.500  &  4.00  &  61.2 &  17.1 & 0.345 \\ \hline
OGLE-33b & &  & & & \\ \hline
764.89532  &  1.895  &  0.70  &  52.0 &  17.2 & 1.797 \\ 
766.87982  &  9.212  &  0.60  &  51.0 &  16.0 & 2.232 \\ 
767.80618  &  $-$63.941  &  1.00  &  48.0 &  15.6 & 1.333 \\ 
769.77114  &  $-$62.742  &  1.00  &  53.0 &  16.9 & 1.293 \\ 
770.85930  &  26.163  &  0.40  &  49.0 &  16.1 & 3.261 \\ 
790.70926  &  22.695  &  0.50  &  52.0 &  16.8 & 2.576 \\ 
797.73001  &  $-$60.852  &  1.10  &  49.0 &  17.1 & 1.117 \\ \hline
\end{tabular}
\end{table}

\begin{table}
\begin{tabular}{r r r r r r}\hline \hline
BJD & RV & depth & FWHM & SNR & $\sigma_{\rm RV}$  \\ 
{[$-2452000\,$d]} & [{\kms}] & [\%] & [{\kms}] & & [{\kms}] \\ \hline 
OGLE-34 & &  & & & \\ \hline
759.79751  &  32.726  &  21.74  &  12.3 &  6.5 & 0.082 \\ 
764.89532  &  88.686  &  20.53  &  11.3 &  6.8 & 0.080 \\ 
766.87982  &  46.630  &  18.25  &  11.6 &  4.9 & 0.119 \\ 
767.80618  &  33.753  &  16.82  &  11.2 &  4.6 & 0.134 \\ 
769.77114  &  53.710  &  20.40  &  12.3 &  5.6 & 0.099 \\ 
770.85930  &  78.537  &  19.60  &  11.3 &  5.3 & 0.103 \\ 
790.70926  &  88.310  &  20.55  &  10.9 &  5.5 & 0.094 \\ 
797.73001  &  96.514  &  19.70  &  12.1 &  5.7 & 0.099 \\ \hline
OGLE-35a & &  & & & \\ \hline
759.79751  &  $-$145.398  &  3.74  &  58.6 &  16.8 & 0.367 \\ 
764.89532  &  $-$134.986  &  3.63  &  57.1 &  17.2 & 0.365 \\ 
766.87982  &  $-$151.252  &  3.59  &  58.0 &  16.0 & 0.399 \\ 
769.77114  &  $-$159.365  &  3.84  &  59.2 &  16.9 & 0.357 \\ 
770.85930  &  42.947  &  3.75  &  56.7 &  16.1 & 0.376 \\ 
790.70926  &  10.277  &  3.47  &  55.0 &  16.8 & 0.383 \\ 
797.73001  &  $-$117.998  &  3.48  &  54.9 &  17.1 & 0.375 \\ \hline
OGLE-35b & &  & & & \\ \hline
759.79751  &  21.308  &  3.78  &  56.7 &  16.8 & 0.357 \\ 
764.89532  &  11.174  &  3.60  &  55.9 &  17.2 & 0.364 \\ 
766.87982  &  27.804  &  3.64  &  56.8 &  16.0 & 0.390 \\ 
769.77114  &  35.511  &  3.78  &  55.9 &  16.9 & 0.353 \\ 
770.85930  &  $-$165.670  &  3.86  &  56.6 &  16.1 & 0.365 \\ 
790.70926  &  $-$133.420  &  3.48  &  55.4 &  16.8 & 0.384 \\ 
797.73001  &  $-$3.016  &  3.42  &  54.9 &  17.1 & 0.382 \\ \hline
OGLE-48 & &  & & & \\ \hline
766.79458  &  $-$0.392  &  0.79  &  203.7 &  12.5 & 4.336 \\ 
767.72839  &  $-$0.798  &  0.89  &  323.6 &  7.8 & 7.774 \\ 
768.84710  &  4.351  &  1.00  &  232.4 &  15.8 & 2.895 \\ 
770.82050  &  6.071  &  1.17  &  383.9 &  8.9 & 5.645 \\ \hline
OGLE-49 & &  & & & \\ \hline
759.72111  &  $-$106.352  &  12.44  &  9.7 &  2.2 & 0.338 \\ 
764.85708  &  $-$106.928  &  13.39  &  7.4 &  2.5 & 0.246 \\ 
766.79458  &  $-$107.230  &  19.76  &  9.5 &  2.7 & 0.174 \\ 
767.72839  &  $-$106.680  &  10.80  &  9.3 &  2.0 & 0.434 \\ 
768.84710  &  $-$106.965  &  25.41  &  9.1 &  4.6 & 0.085 \\ 
769.85598  &  $-$107.327  &  17.46  &  9.4 &  2.6 & 0.204 \\ 
770.82050  &  $-$106.830  &  16.52  &  9.7 &  2.6 & 0.218 \\ 
797.76880  &  $-$106.970  &  20.26  &  9.4 &  3.1 & 0.150 \\ \hline
\end{tabular}
\end{table}

\begin{table}
\begin{tabular}{r r r r r r}\hline \hline
BJD & RV & depth & FWHM & SNR & $\sigma_{\rm RV}$  \\ 
{[$-2452000\,$d]} & [{\kms}] & [\%] & [{\kms}] & & [{\kms}] \\ \hline 
OGLE-55 & &  & & & \\ \hline
759.79751  &  $-$12.937  &  5.62  &  49.2 &  7.0 & 0.536 \\ 
764.89532  &  5.990  &  5.55  &  50.7 &  7.3 & 0.528 \\ 
766.87982  &  24.803  &  4.20  &  48.4 &  5.8 & 0.857 \\ 
767.80618  &  20.569  &  5.18  &  53.1 &  5.2 & 0.812 \\ 
769.77114  &  9.084  &  5.73  &  53.1 &  6.0 & 0.637 \\ 
770.85930  &  26.478  &  5.38  &  46.4 &  6.4 & 0.595 \\ 
790.70926  &  $-$12.510  &  4.66  &  43.5 &  5.4 & 0.787 \\ 
797.73001  &  $-$21.156  &  5.47  &  49.1 &  7.5 & 0.514 \\ \hline
OGLE-56 & &  & & & \\ \hline
759.79751  &  $-$48.284  &  30.47  &  10.1 &  12.2 & 0.043 \\ 
764.89532  &  $-$48.528  &  30.83  &  10.2 &  11.6 & 0.044 \\ 
766.87982  &  $-$48.241  &  29.47  &  10.7 &  9.2 & 0.050 \\ 
767.80618  &  $-$48.279  &  27.69  &  10.5 &  8.0 & 0.056 \\ 
769.77114  &  $-$48.516  &  29.69  &  10.3 &  10.3 & 0.047 \\ 
770.85930  &  $-$48.485  &  29.44  &  10.5 &  9.8 & 0.049 \\ 
790.70926  &  $-$48.306  &  28.20  &  10.3 &  9.0 & 0.052 \\ 
797.73001  &  $-$48.518  &  29.11  &  10.1 &  9.9 & 0.048 \\ 
809.76233  &  $-$48.550  &  36.90  &  8.1 &  2.7 & 0.087 \\ 
810.59822  &  $-$48.126  &  40.30  &  8.6 &  2.3 & 0.094 \\ 
810.83456  &  $-$48.440  &  37.40  &  9.1 &  2.3 & 0.107 \\ 
811.63656  &  $-$48.011  &  40.10  &  8.2 &  3.8 & 0.057 \\ 
811.83617  &  $-$48.212  &  39.50  &  8.1 &  3.0 & 0.073 \\ \hline
OGLE-58 & &  & & & \\ \hline
759.75994  &  51.016  &  11.48  &  13.1 &  15.6 & 0.070 \\ 
764.81842  &  51.091  &  11.69  &  12.6 &  13.8 & 0.075 \\ 
766.84092  &  51.115  &  11.54  &  13.0 &  14.4 & 0.074 \\ 
767.76871  &  51.200  &  11.04  &  12.6 &  10.5 & 0.098 \\ 
768.76694  &  51.169  &  11.70  &  12.7 &  16.5 & 0.066 \\ 
769.81610  &  51.051  &  11.60  &  13.0 &  15.1 & 0.071 \\ 
770.90035  &  50.996  &  11.35  &  13.1 &  13.8 & 0.078 \\ 
791.72546  &  51.004  &  11.38  &  12.9 &  15.3 & 0.071 \\ \hline
OGLE-59a & &  & & & \\ \hline
759.75994  &  9.765  &  3.32  &  49.4 &  13.3 & 0.479 \\ 
764.81842  &  8.670  &  3.30  &  49.5 &  10.8 & 0.593 \\ 
766.84092  &  9.714  &  3.30  &  46.6 &  12.0 & 0.518 \\ 
767.76871  &  8.835  &  3.30  &  49.5 &  8.0 & 0.800 \\ 
768.76694  &  8.890  &  3.43  &  51.3 &  14.1 & 0.446 \\ 
769.81610  &  8.119  &  3.43  &  49.6 &  12.9 & 0.479 \\ 
770.90035  &  9.113  &  3.30  &  49.5 &  10.8 & 0.593 \\ 
791.72546  &  9.363  &  3.30  &  49.5 &  12.4 & 0.517 \\ \hline
OGLE-59b & &  & & & \\ \hline
759.75994  &  61.728  &  2.28  &  27.0 &  13.3 & 0.515 \\ 
764.81842  &  3.619  &  3.11  &  36.7 &  10.8 & 0.542 \\ 
766.84092  &  $-$44.414  &  2.82  &  28.6 &  12.0 & 0.475 \\ 
767.76871  &  $-$2.520  &  2.46  &  27.2 &  8.0 & 0.796 \\ 
768.76694  &  60.252  &  2.63  &  25.3 &  14.1 & 0.408 \\ 
769.81610  &  $-$44.422  &  2.89  &  26.4 &  12.9 & 0.415 \\ 
770.90035  &  13.946  &  2.92  &  34.8 &  10.8 & 0.562 \\ 
791.72546  &  $-$1.157  &  3.70  &  35.5 &  12.4 & 0.391 \\ \hline
\end{tabular}
\end{table}

\subsection{Rotation velocities}

For each object, the eight observed cross-correlation function were shifted by the 
observed radial velocity and co-added to give a combined CCF of higher signal-to-noise ratio.
Rotationally broadened line profiles were convolved with a Gaussian instrumental profile 
depending of the instrument and correlation mask: $\sigma=3.0$ {\kms} for UVES and $4.0$ {\kms}
for FLAMES. The instrumental profile was determined with HD162907 for UVES and the 
combined spectrum of OGLE-TR-19 and OGLE-TR-49 for FLAMES. We also checked 
the instrumental profile on both spectrographs with the ThAr spectrum. The profiles were 
fitted to the CCF to determine the projected rotation velocity {\vsini} of the target objects.
The result is displayed in Table~\ref{tablespectro}. 
A  quadratic limb-darkening with coefficients  $u1\! + \! u2\! =\! 0.6$ was assumed. 
The computations of Barban et al. (\cite{barban}) find that such a coefficient is a suitable 
approximation for a wide range of spectral types in wavelengths corresponding to the V filter. 

For close binaries, with rotation periods of the order of a few days, we expect that the 
rotation axis is aligned with the orbital axis, the orbit is circularized and the system 
is tidally locked (e.g., Levato \cite{levato}, Hut \cite{hut}, Melo et al. \cite{melo}). 
For known close binaries, the alignment of the axes and the tidal locking are observed to be 
effective even before orbital circularization. 
It can therefore be expected that in cases of a massive transiting companion with a short period, 
the system is tidally locked and {\vsini} is large. In that case, $P_{rot}=P_{transit}$ and the 
rotational velocity is directly related to the radius of the primary. Rotation velocities 
observed in our sample are 
generally compatible with the hypothesis of tidal locking. In these cases {\vsini} provides a 
measurement of $R$ with an estimated accuracy of a few percent.
The uncertainty in the determination of {\vsini} was estimated by computing values 
on each of the individual CCF and calculating the dispersion of these values. In most 
cases this "formal" uncertainty is very small, and the dominant source of error is 
actually the adopted value of the limb darkening coefficient (see Section~\ref{discu}). 
For the smallest rotational velocities ({\vsini} $ < 5$~{\kms}), the dominant 
uncertainty becomes the adopted value of the instrumental broadening and the 
stellar micro-turbulence parameter. In order to take into account such systematic 
uncertainties we fix afterward the lowest uncertainty of {\vsini} to 1~{\kms}.  

\subsection{Stellar spectroscopic parameters}

For the slowly-rotating stars in our sample, the stellar parameters
(temperatures, gravities and metallicities) were obtained from an analysis 
of a set of \ion{Fe}{i} and \ion{Fe}{ii} lines, following the procedure used 
in Santos et al. (\cite{santos04}). Line equivalent widths were derived 
using IRAF\footnote{distributed by NOAO, AURA, Inc., under contract with the NSF, USA.}, 
and the abundances were obtained using 
a revised version of the code MOOG (Sneden \cite{sneden}), and a grid of 
Kurucz (\cite{kurucz}) atmospheres. 

The final parameters have errors of the order of 200\,K in
T$_{\mathrm{eff}}$, 0.40 in $\log{g}$, and 0.20\,dex in [Fe/H] (see
Table~\ref{tablespectro}). The precision of the derived atmospheric parameters 
is mostly affected by relatively low S/N of the combined spectra (30-50), 
together with some possible contamination coming from the ThAr spectrum. 
Furthermore, blends with the spectrum of a low mass stellar companion or a 
background star may also affect the determination of the stellar parameters.

For the fast rotating stars in our sample ({\vsini} $\geq 20$~{\kms}),
the method described in Santos et al. (\cite{santos04}) is not applicable,
given that the measurement of individual equivalent widths is
not accurate enough due to line-blending. In these cases we
have adopted a simpler approach, and have determined very approximate
effective temperatures for the stars by visual comparison of the combined spectra 
with synthetic spectra convolved with a rotational profile to take into account the
 projected rotational velocity.

Stellar spectroscopic parameters are given in Table~\ref{tablespectro}. 
The lines of OGLE-TR-48 are too rotationally broadened for a 
spectral type estimation. Our $\log g$ and $[Fe/H]$ estimates are all very uncertain, 
they simply indicate that the target objects are dwarfs and have solar or above-solar 
metallicities.\\

\begin{table}
\caption{Parameters from the spectroscopic analysis. Rotation velocities {\vsini} are 
computed from the analysis of the CCF. $T_{eff}$, $\log g$, [Fe/H]: temperature, gravity 
and metallicity are computed from the analysis of the spectral lines in case of low \vsini. 
For high rotation, $T_{eff}$ was estimated roughly by comparison with synthetic spectrum. 
Spectral type of binaries (identified with label $a$ and $b$) which include two moving 
components in the spectra was not possible to determine.}
\label{tablespectro}
\begin{tabular}{r r r r r} \hline \hline
Name & \vsini & $T_{eff}$ & $\log g$ & [Fe/H]\\ 
 & [\kms] & [K] &  & \\ \hline 
5 & 87.8 $\pm$ 1.8 & 6700$\pm$500 & - & - \\
6 & 22.6 $\pm$ 0.3 & 5700$\pm$500 & - & - \\
7 & 30.7 $\pm$ 0.2 & 6500$\pm$500 & - & - \\
8a & 11.7 $\pm$ 0.3 & - & - &  -\\
8b & 11.3 $\pm$ 0.4 & - & - & - \\
10 & 7.7 $\pm$ 0.1 & 6220$\pm$140 & 4.70$\pm$0.34 & 0.39$\pm$0.14 \\
12 & 7.7 $\pm$ 0.2 & 6440$\pm$300 & 4.90$\pm$0.35 & 0.30$\pm$0.26 \\
17 & $<$5 & 5870$\pm$190 & 4.80$\pm$0.45 & $-$0.06$\pm$0.21 \\
18 & 44.3 $\pm$ 0.9 & 6500$\pm$500 & - & - \\
19 & 6.0 $\pm$ 0.2  & 5250$\pm$300 & 4.0$\pm$0.50 & $-$0.10$\pm$0.30 \\
33a & 45.3 $\pm$ 0.5 & 6700$\pm$500 & - & - \\
33b & $\sim$ 32 & - & - & - \\
34 & 6.3 $\pm$ 0.3 & 6520$\pm$340 & 4.50$\pm$0.48 & 0.32$\pm$0.31 \\
35a & 34.3 $\pm$ 1.1 & - & - & - \\
35b & 34.8 $\pm$ 0.8 & - & - & - \\
48 & $>$100  & - & - & - \\
49 & $<$5  & 5000$\pm$500 & - & - \\
55 & 30.5 $\pm$ 0.6 & 6000$\pm$500 & - & - \\
56 & $<$5 & 5970$\pm$150 & 4.20$\pm$0.38 & 0.17$\pm$0.19 \\ 
58 & 14.6 $\pm$ 0.1  & 6500$\pm$500 & - & - \\
59a & 34.1 $\pm$ 0.7  & - & - & - \\
59b & 20.9 $\pm$ 1.1  & - & - & - \\ \hline
\end{tabular}
\end{table}

A spectral classification of 7 of our targets (OGLE-TR-5, 6, 8, 10, 12, 19 and 35) 
was previously done by Dreizler et al. (\cite{dreizler1}). Our stellar spectroscopic 
parameters determination is in agreement with their result.
  
\section{Light curve analysis}

\subsection{Existence and periodicity of transit signal}
\label{existence}

When the amplitude of the radial velocity variation is small, which may indicate 
the presence of a planet or a blend with a background eclipsing binary, one must first 
consider the possibility that the detected photometric signal is not a bona fide 
transit/eclipse, or that the period of the signal is incorrect. The OGLE photometric 
data is subject to systematic intra-night drifts in calibration to the level of the 
hundredth of magnitude, similar to the depth of the smallest detected transit signal. 
Given the fact that the transit candidates were detected among about 60'000 light curves, 
it is not impossible that some of them are artefacts. Either the whole detection is 
spurious or, more likely, one of the detected transit is an artifacts. If there are 
only two detected transits, this would imply that the periodicity of the event is unknown, 
and therefore that no information can be derived from the absence of radial velocity 
variations.

\begin{figure}
\resizebox{8.5cm}{!}{\includegraphics{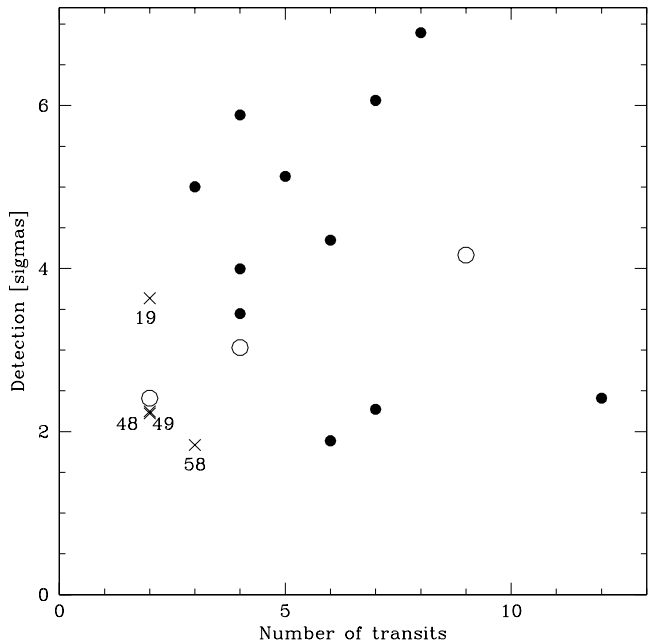}}
\caption{Number of detected transits vs. $d/\sigma$, where $d$ is the depth of the 
transit signal and $\sigma$ the photometric uncertainty. Open symbols indicate objects for which 
the period had to be modified according to the radial velocity data (OGLE-TR-12, 17 and 59), 
crosses the objects for which no radial velocity variation was detected.}
\label{confidence}
\end{figure}

To quantify the reliability of the transit detection, we divided the value of the transit 
depth by its uncertainty, yielding a "confidence factor" for the existence of the transit, 
and plotted this  factor as a function of the number of transits (see Fig.~\ref{confidence}). 
Objects with only two or three detected transits and  a low confidence factor are more 
likely to be artefacts. This is the case of objects OGLE-TR-19, 48, 49 and 58 which 
are possible spurious transit detections and are discussed individually below 
(Section~\ref{false} and Fig.~\ref{doubtful}). 

In cases when the radial velocity data does not obviously confirm the period of the 
transit candidates, we also studied the light curve to see if the transit signal were 
compatible with other periods.  For instance, in a two-transit case, several divisors of the 
interval between the two transits 
can be possible to put the data in phase. For OGLE-TR-12, 17 and 59, another 
period than the one given by Udalski et al. (\cite{udalski1}, \cite{udalski2}) was found to phase 
the light curve and radial velocity data perfectly. In case of equal mass double-line eclipsing 
binaries (OGLE-TR-8 and 35), the right period is twice the OGLE period due to the fact that both 
transits and anti-transits are present in the photometric curve.\\

\subsection{Sinusoidal variations in the light curve}
\label{s2}

As repeated in Sirko \& Paczynski (\cite{sirko}), close binaries can induce variability to the 
mmag level in the light curve in phase with the transit signal period. If the light of 
the secondary is not negligible compared to the primary, an anti-transit signal can be 
visible. Even in the absence of anti-transit, the ellipsoidal deformation of the primary 
under the gravitational influence of the secondary causes sinusoidal variations in the light curve with 
double the phase of the orbital period. Such sinusoidal signals were fitted to the OGLE 
transit candidate light curve by Sirko \& Paczinski (\cite{sirko}). 

We have repeated their procedure and find very close results except for the objects 
for which the period had to be modified according to the radial velocity data (OGLE-TR-12, 
17, and 59) and for the SB2 (OGLE-TR-8 and 35). Such procedure clearly indicates that 
the OGLE-TR-5, 8, 18 and 35 have a massive companion in the stellar mass range.    

The periodic sinusoidal signals in the light curves were subtracted when significant from the 
data before the analysis of the transit shape.

\subsection{Analysis of the transit shape}
\label{agol}

The depth, width and general shape of the transit signal depend on a combination of 
physical variables, mainly the radius ratio $\overline{r}$, the primary radius $R$ and the impact 
parameter $b$ (or, equivalently for circular orbits, the angle $i$ of the normal of the 
orbital plane with the line-of-sight) and the orbital eccentricity. It is also more weakly 
dependent on the total mass $(m+M)$ -- via the orbital period and semi-major axis for a 
Keplerian orbit -- and the limb darkening coefficients. The parameter $\overline{r}$ is mainly 
constrained by the transit depth, 
$R \cdot (m+M)^{-1/3}$ by its duration, and $b$ by its shape. We assumed that all orbits were 
circular ($e=0$). Low-period binaries below $P\sim 10$ days are observed to have circularized 
orbits (Levato \cite{levato}), and all objects in our sample have lower periods except OGLE-TR-17, which 
indeed shows indications of a small eccentricity in the radial velocity curve (See 
Section~\ref{ogle17}). In all other cases of large-amplitude variations the radial velocity 
residuals show no significant variations from a circular orbit. 

The light curve were fitted by non-linear least square fitting with analytic transit curves 
computed according to Mandel \& Agol (\cite{mandel}), using a 
quadratic limb darkening model with $u1\! + \! u2\! =\! 0.3$. Notice that this is 
different from the coefficients used for the determination of the rotational velocity, 
because the wavelengths are different. The OGLE data was 
obtained with an $I$ filter while the spectra are centered on the visible. The fitted 
parameters were $\overline{r}$, $V_T/R$ and $b$, where $V_T$ is the transversal orbital velocity 
at the time of the transit. 

Broadly, there are two kinds of cases for the transit shape. Either the transit signal 
is broad and flat and $b$ has a firm higher bound -- a central transit -- and in that 
case $\overline{r}$ and $V_T/R$ are very well constrained by the depth and duration of the transit. Or the 
V-shaped or indefinite signal shape allows for high values of $b$ -- a grazing 
transit -- and in that case $\overline{r}$ and $V_T/R$ are correlated with $b$ and cannot 
be well determined independently. In real terms, it means that the signal comes either 
from a small body transiting rapidly across the primary, or a larger body partially 
obscuring the primary in a slower grazing eclipse. 
For illustration, Fig.~{\ref{fittransit} shows the fit of the transits of all our candidates.

\begin{figure*}
\resizebox{17cm}{!}{\includegraphics{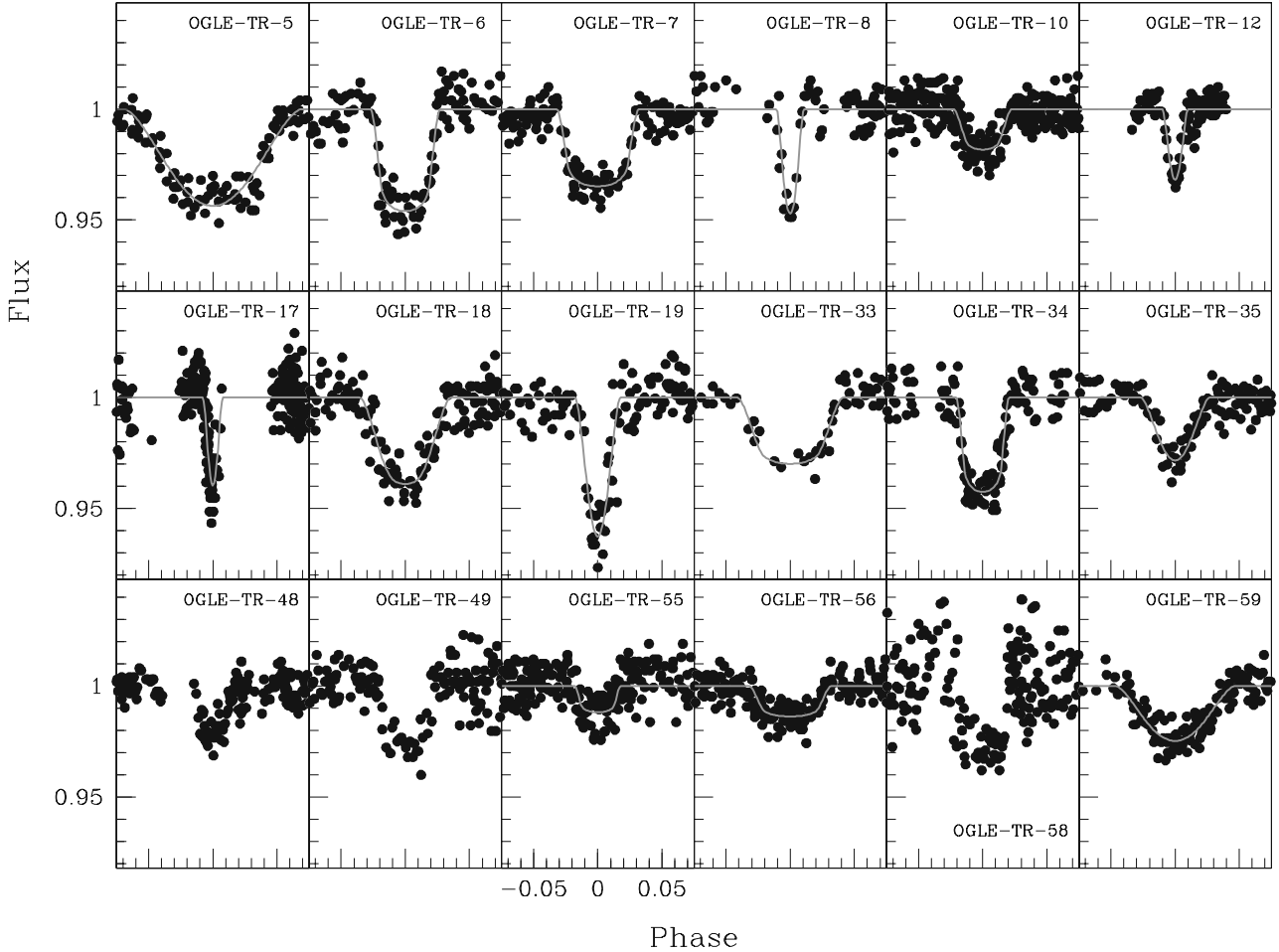}}
\caption{Phase-folded light curve and best-fit transit curve for all the followed OGLE 
candidates.}
\label{fittransit}
\end{figure*}

The OGLE photometric data is subject to strongly covariant noise. For many objects the systematic 
drift during the nights -- especially near the beginning and end of the night -- are of the same 
order as the random noise or even larger (see lowest panel on Fig.~\ref{doubtful}). Therefore, all the data points cannot 
be considered as independent estimators on the fitted curve and an error analysis from a chi-square 
distribution will significantly underestimate the uncertainties.

To compute realistic uncertainties on the values of the transit parameters, we used a technique 
based on the permutation of the residuals. We compute the residuals by removing the best-fit 
solution for the transit shape, then exchange the residuals of one night with another randomly 
chosen night. The fitting procedure is then repeated on the resulting curve with the shuffled 
residuals, leaving the period as a free parameter. Many Monte Carlo realizations are then 
carried out to estimate the dispersion of the fitted parameters.

This permutation procedure has the advantage of ``letting the data speak by itself'' and automatically 
incorporates the real characteristics of both the random noise and the systematic intra-night drifts. 
Its use is especially relevant in objects where the transit was covered in a low number of nights. 
In that case, intra-night systematic drifts can significantly alter the transit shape, and a 
$\chi^2$ analysis will yield much too low uncertainties. Figure \ref{permutation} gives an example 
of the residual permutation procedure on OGLE-TR-12.

\begin{figure}
\resizebox{8.5cm}{!}{\includegraphics{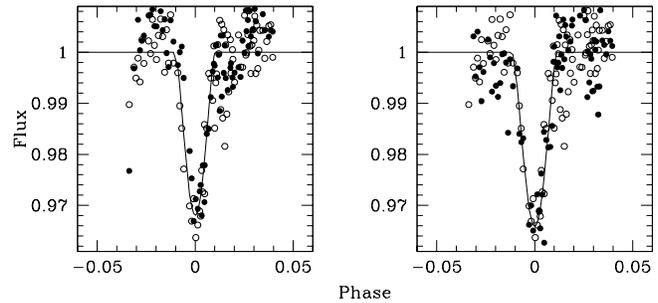}}
\caption{Illustration (for OGLE-TR-12) of the residual permutation method for the estimation 
of the uncertainties. On both panels, the open symbols indicate the original data, and the 
dots indicate two realizations (among a total of 100 realizations) of the data with 
residuals permuted between different nights. The lines indicate the best-fit curve on 
the permitted data. The resulting $\overline{r}$ is 0.254 and 0.292 respectively. Because the transit 
coverage is constituted of only two nights, systematic trends in the residuals have a large 
effect on the resulting parameters. In the case of OGLE-TR-12, this error estimate leads 
to much higher 1-sigma interval than a $\chi^2$ analysis.}
\label{permutation}
\end{figure}

\begin{table*}
\caption{Parameters from the transit light curve fit: $\overline{r}$ 
radius ratio of the primary and secondary bodies, $V_T/R$ transit velocity in units of the 
primary radius, $b$ impact parameter, $P$ revisited period according to the radial velocity 
measurements. We deliberately do not provide the results for the 4 unsolved cases suspected 
to be false positives OGLE-TR-19, 48, 49 and 58. For the two cases of SB2, stellar parameters 
were deduced from the spectroscopic orbits and the rotational velocities.}
\label{tabletransit}
\begin{tabular}{r r r r r r}\hline \hline
Name & $\overline{r}$ & $V_T/R$ & b & $[b_{down}-b_{up}]$ & P \\ 
 &  & [days$^{-1}$] &  & & [days]\\  \hline 
5 &  0.189 $\pm$ 0.008 & 22.70 $\pm$ 1.9 & 0.56 & [0.47-0.65] & 0.80827  \\
6 &  0.198 $\pm$ 0.006 & 9.76 $\pm$ 0.44 & 0.20 & [0.00-0.21] & 4.54881  \\
7 & 0.172 $\pm$ 0.006 & 13.58 $\pm$ 0.50 & 0.02 & [0.00-0.09] & 2.71782  \\
8 & -  & -  & 1.65 & [1.62-1.68] & 5.43284  \\
10 & 0.129 $\pm$ 0.007 & 16.80 $\pm$ 1.22 & 0.45 & [0.00-0.46] & 3.10140  \\
12 & 0.299 [0.26-0.74]  & 8.44 $\pm$ 0.83 & 1.02 & [0.89-1.32] & 8.65725  \\
17 & 0.245 [0.22-0.63] & 8.27 $\pm$ 0.56 & 0.90 & [0.72-1.28] & 13.87370  \\
18 & 0.200 $\pm$ 0.020 & 11.78 $\pm$ 0.83 & 0.77 & [0.73-0.87] & 2.22803 \\
19 & - & -  & - & - & - \\
33 & 0.160 $\pm$ 0.015 & 12.20 $\pm$ 1.25 & 0.68 & [0.59-0.77] & 1.95326  \\
34 & 0.188 $\pm$ 0.002 & 6.83 $\pm$ 0.40 & 0.00 & [0.00-0.02] & 8.57633  \\
35 & -  & -  & 1.68 & [1.65-1.73] & 2.51957  \\
48 & -  & -  & - & - & - \\
49 & -  & -  & - & - & - \\
55 & 0.109 [0.107-0.150] & 16.45 [9.21-18.72] & 0.60 & [0.00-0.85] & 3.18454  \\
56 & 0.114 $\pm$ 0.004 & 22.84 [22.43-27.27] & 0.69 & [0.50-0.69] & 1.21192  \\
58 & - & -  & - & - & - \\
59 & 0.257 $\pm$ 0.089 & 10.238 $\pm$ 0.632 & 0.99 & [0.82-1.31] & 2.99432 \\ \hline
\end{tabular}
\end{table*}

\subsection{Synthesis of the spectroscopic and photometric constraints}
\label{synth}

The different constraints are combined by chi-square minimization to obtain an estimate of the 
physical characteristics of the two bodies involved in the transit. In most cases, constraints 
overlap and allow one or several coherence checks between the different lines of inquiry.

The measured rotation velocity {\vsini} (in $km\,s^{-1}$) is related to the primary radius 
(in solar unit) and the period (in days) through:
\begin{equation}
v\,\sin i  = 50.6 \cdot \frac{R}{P} \cdot {\sin i}\,.
\end{equation}

For a circular orbit, the semi-amplitude of the radial velocity variation (in {\kms}) is related 
to the masses (in solar unit) and period (in days) through:
\begin{equation}
K=214 \cdot \frac{m}{(m+M)^{2/3}} \cdot P^{-1/3}\,.
\end{equation}

The impact parameter $b$ and transit velocity per primary radius $V_T$/R are related to the orbital 
and physical parameters $R$, $m$, $M$ (in solar unit), $P$ (in days) and $i$, for a circular 
Keplerian orbit, in the following way:
\begin{equation}
b= \frac{a \cdot \cos i}{R}\,,  
\end{equation}
\begin{equation}
V_T/R=2 \pi \cdot  \frac{a}{P \cdot R}\,, 
\end{equation}
with $a$ (in solar radius) given by the third Kepler's law: 
\[
a= 4.20 \cdot P^{2/3} \cdot (m+M)^{1/3}\,.
\]
The radius ratio is obviously related to the stellar radius $R$ and companion radius $r$:
\begin{equation}
\overline{r}=r \cdot R^{-1}\,. 
\end{equation}

Finally, the spectroscopic determinations of the temperature, gravity and metallicity of the 
primary provides independent constraints on $R$ and $M$ via the stellar evolution models. 
We used the relation between physical and observable parameters provided by the models of 
Girardi et al. (2002).

\begin{equation}
(\log T_{eff}, \log g, [Fe/H], R)= f(M, age, Z)\,.
\end{equation}

The constraints of Equations 1 to 6 were combined by chi-square minimization. In most cases, $R$ is 
precisely determined from the rotation velocity. Then $M$ is constrained on the one hand by the 
spectroscopic parameters applied by the models, on the other hand by the $V_T/R$ factor in the fit 
and the transit curve. Parameter $r$ then proceeds from $R$ and $\overline{r}$, and $m$ from $M$ and $K$. Parameter 
$i$ is mainly derived from $b$. A low value of the minimum $\chi^2$ ensures that the different constraints 
are coherent. This procedure yields very satisfactory accuracies on all the derived parameters. 
Note that the resulting $r$ and $m$ are only weakly dependent on the accuracy of the stellar 
evolution models, because $R$ is primarily determined from the rotation velocity and $M$ enters 
the determination of $m$ with a 2/3 exponent.

There are two cases for which the accuracy on the final parameters is lower: 1) In cases when 
high values of $b$ ($b\sim 1$) are compatible with the light curve data, 
there is some degeneracy between the impact parameter, the duration of the eclipse and the 
radius ratio. In that case the upper error bar on $r$ can be large because the light curve 
data is compatible with a grazing eclipse by a larger object. This is the case of OGLE-TR-12, 
17 and 55. 2) If the primary is not in synchronous rotation, then $R$ and $M$ are much more 
weakly constrained by equations 5 and 6. In that case the influence of the adequacy of the 
stellar evolution models becomes more important. This is the case of OGLE-TR-34.

\section{Results}
\label{results}

In this section we present our results of Doppler follow-up and light curve analysis. 
Figure \ref{sb1}, \ref{sb2}, \ref{sb3}, 
and \ref{planet} show the radial velocity data phased with the period from Udalski et al. 
(\cite{udalski1}, \cite{udalski2}), or, in the case of OGLE-TR 8, 12, 17, 35, and 59, with the modified 
period obtained from our analysis. If the radial velocity variations are caused by the transiting 
objects, then phase $\phi=0$ must correspond to the passage of the curve at center-of-mass velocity 
with decreasing velocity, which provides a further constraint.
Results of the fit of the transit shape are summarized in Table~\ref{tabletransit}.
Notice that we deliberately do not provide the results for the 4 unsolved cases suspected 
to be false positives. For the two cases of grazing eclipsing binaries, the stellar parameters 
were not deduced from the light curve (except the impact parameter $b$) but from the 
spectroscopic orbits and the rotational velocities.

We distinguish 5 classes of objects, 
the low mass star transiting companions, the grazing eclipsing binaries, 
the triple systems, the planetary candidates, and 
the unsolved cases. For the majority of objects, we used the photometric ephemeris given by 
Udalski et al. (\cite{udalski1}, \cite{udalski2}) and updated ephemeris available from the OGLE 
website \footnote{http://bulge.princeton.edu/$\sim$ogle/ogle3/transits/transit\_news.html}. 
We fixed the transit epoch $T0$ and fit the phased radial velocity with a circular orbit. 
In this way, we determined an updated or corrected period $P$, the velocity offset $V0$, 
and the velocity semi-amplitude $K$. Each classes of objects are described in the following 
subsection and the derived masses and radii are presented and discussed in section~\ref{discu}.

\subsection{low mass star transiting companions}

\begin{table*}
\caption{Orbital parameters of the low mass star transiting companions, the grazing 
eclipsing binaries and the triple systems. For the triple systems, the component $a$ correspond 
here to the contaminating third body with fixed parameters and component $b$ to the primary of 
an eclipsing binary with fitted parameters. $^1$ Residuals of OGLE-TR-17 without and with 
eccentricity (e=0.074). $^{2}$ This value corresponds to the anti-transit epoch, the revised value is 74.65125.}
\label{tableKsb1}
\label{tableKsb2}
\label{tableKsb3}
\begin{tabular}{r r r r r r r}\hline \hline
Name & P$_{OGLE}$ & T0$_{OGLE}$ & P & K & V0 & O-C \\ 
 & [days] & $-$2452000 & [days] & [{\kms}]& [{\kms}] & [{\kms}]\\  \hline 
5 & 0.8082 & 60.47118 & 0.808271 & 49.90 $\pm$ 2.0 & 16.1 $\pm$ 1.4 & 3.5 \\ 
6 & 4.5487 & 61.05651 & 4.54881 & 32.09 $\pm$ 0.15 & 15.33 $\pm$ 0.10 & 0.21 \\
7 & 2.7179 & 61.42566 & 2.71824 & 31.39 $\pm$ 0.22 & $-$10.94 $\pm$ 0.16 & 0.37 \\
12 & 5.7721 & 61.53515 & 8.65725 & 34.862 $\pm$ 0.025 & 24.255 $\pm$ 0.019  & 0.083 \\
17 & 2.3171 & 62.35748 & 13.8737 & 31.99 $\pm$ 0.19 & $-$34.36 $\pm$ 0.12 & 0.894/0.072$^1$ \\
18 & 2.2280 & 61.07501 & 2.228025 & 46.87 $\pm$ 1.48 & $-$46.75 $\pm$ 1.03 & 2.08 \\
34 & 8.5810 & 62.74970 & 8.57633 & 32.68 $\pm$ 0.27 & 65.50 $\pm$ 0.20 & 0.426 \\
55 & 3.18456 & 77.05049 & 3.184543 & 28.63 $\pm$ 0.32 & 5.06 $\pm$ 0.19 & 0.439 \\ \hline
8a & 2.7152 & 61.01604 & 5.432842 & 79.13 $\pm$ 0.14 & $-$3.71 $\pm$ 0.07 & 0.260 \\
8b & 2.7152 & 61.01604 & 5.432842 & 82.10 $\pm$ 0.15 & $-$3.71 $\pm$ 0.07 & 0.299 \\
35a & 1.2599 & 60.98942 & 2.519569 & 104.57 $\pm$ 0.38 & $-$61.53 $\pm$ 0.22 & 0.521 \\ 
35b & 1.2599 & 60.98942 & 2.519569 & 104.21 $\pm$ 0.37 & $-$61.53 $\pm$ 0.22 & 0.611 \\ \hline
33a & 1.9533 & 60.54289 & - &  0.0 & $-$28.5  & - \\
33b & 1.9533 & 60.54289 & 1.95327 &  59.8 $\pm$ 2.1 & $-$29.8 $\pm$ 1.3 & 3.2 \\
59a & 1.49709 & 73.15416 & - &  0.0 & 9.05  & 0.495 \\ 
59b & 1.49709 & 73.15416$^2$ & 2.9943224 &  58.68 $\pm$ 0.31 & 6.01 $\pm$ 0.22 & 0.524 \\ \hline
\end{tabular}
\end{table*}

\begin{figure*}
\resizebox{8.5cm}{!}{\includegraphics{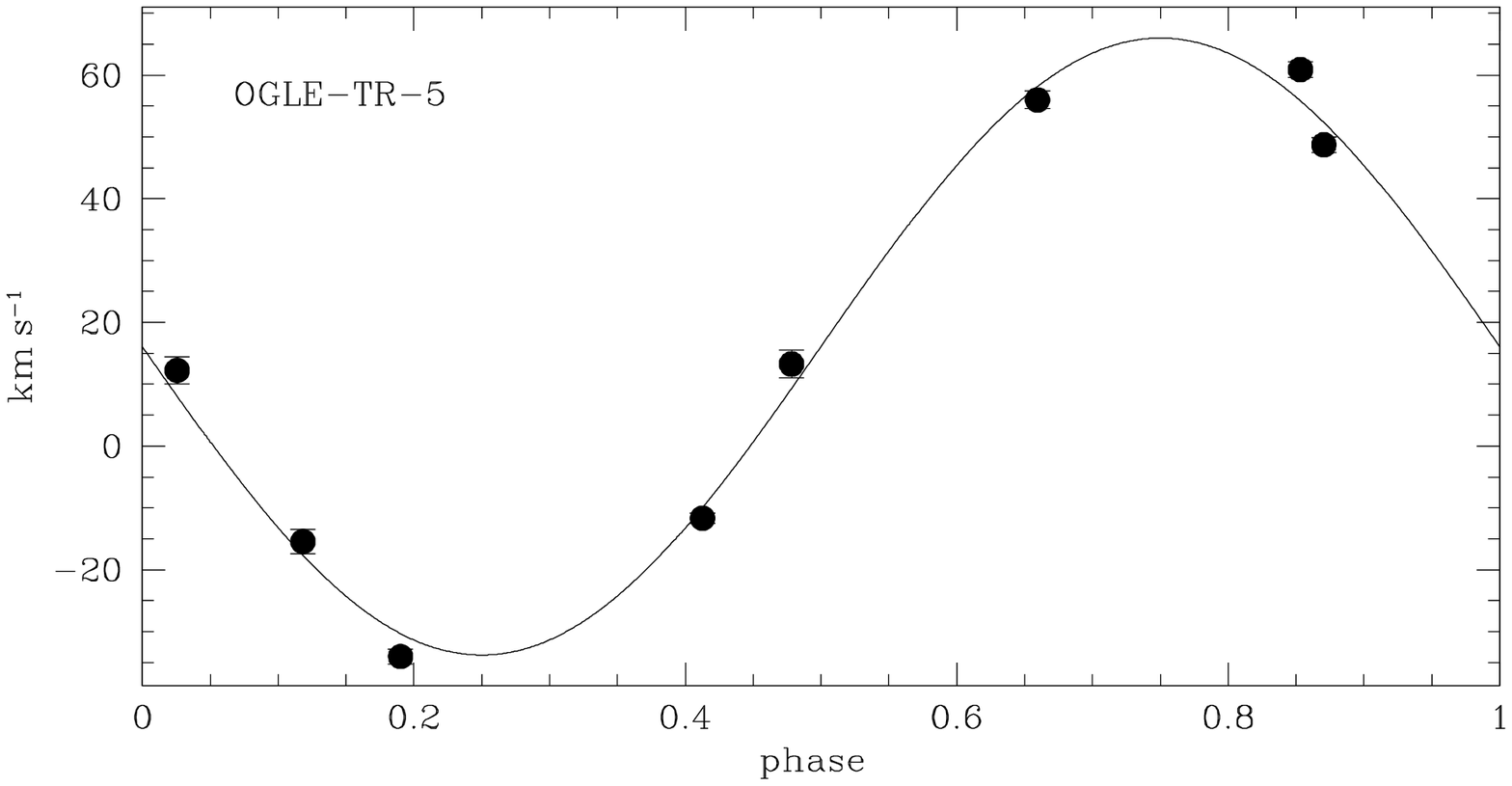}}
\resizebox{8.5cm}{!}{\includegraphics{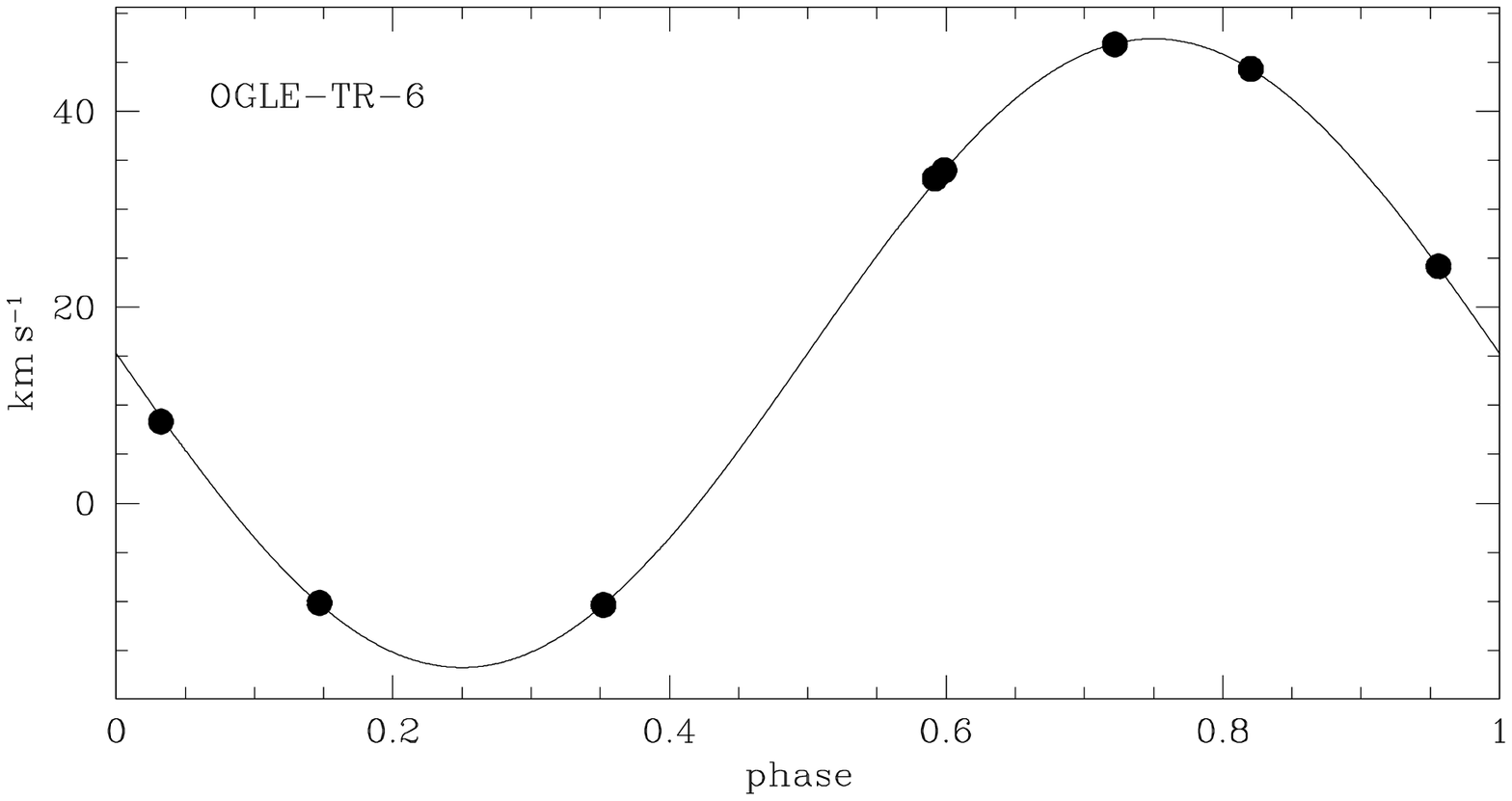}}

\resizebox{8.5cm}{!}{\includegraphics{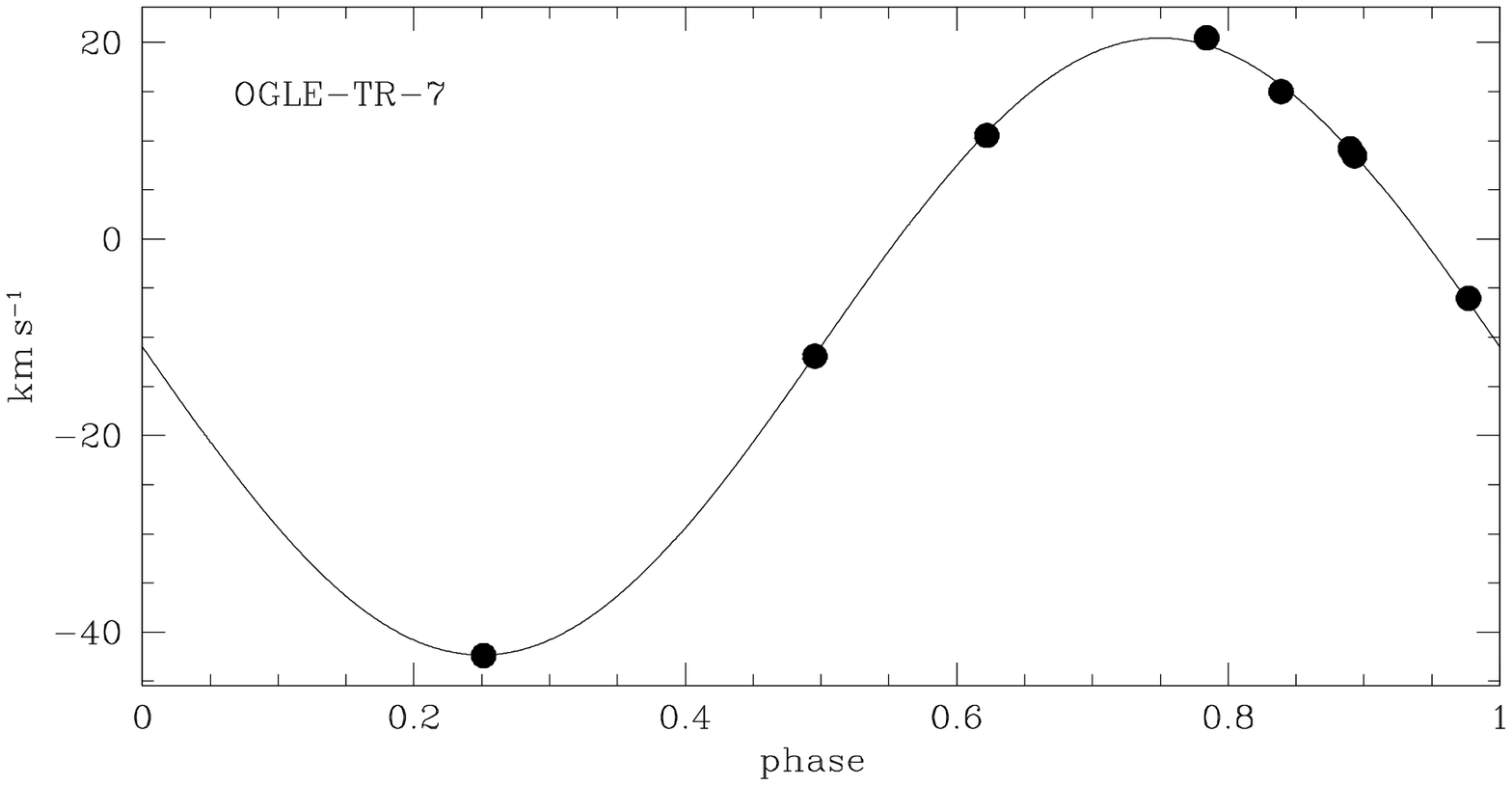}}
\resizebox{8.5cm}{!}{\includegraphics{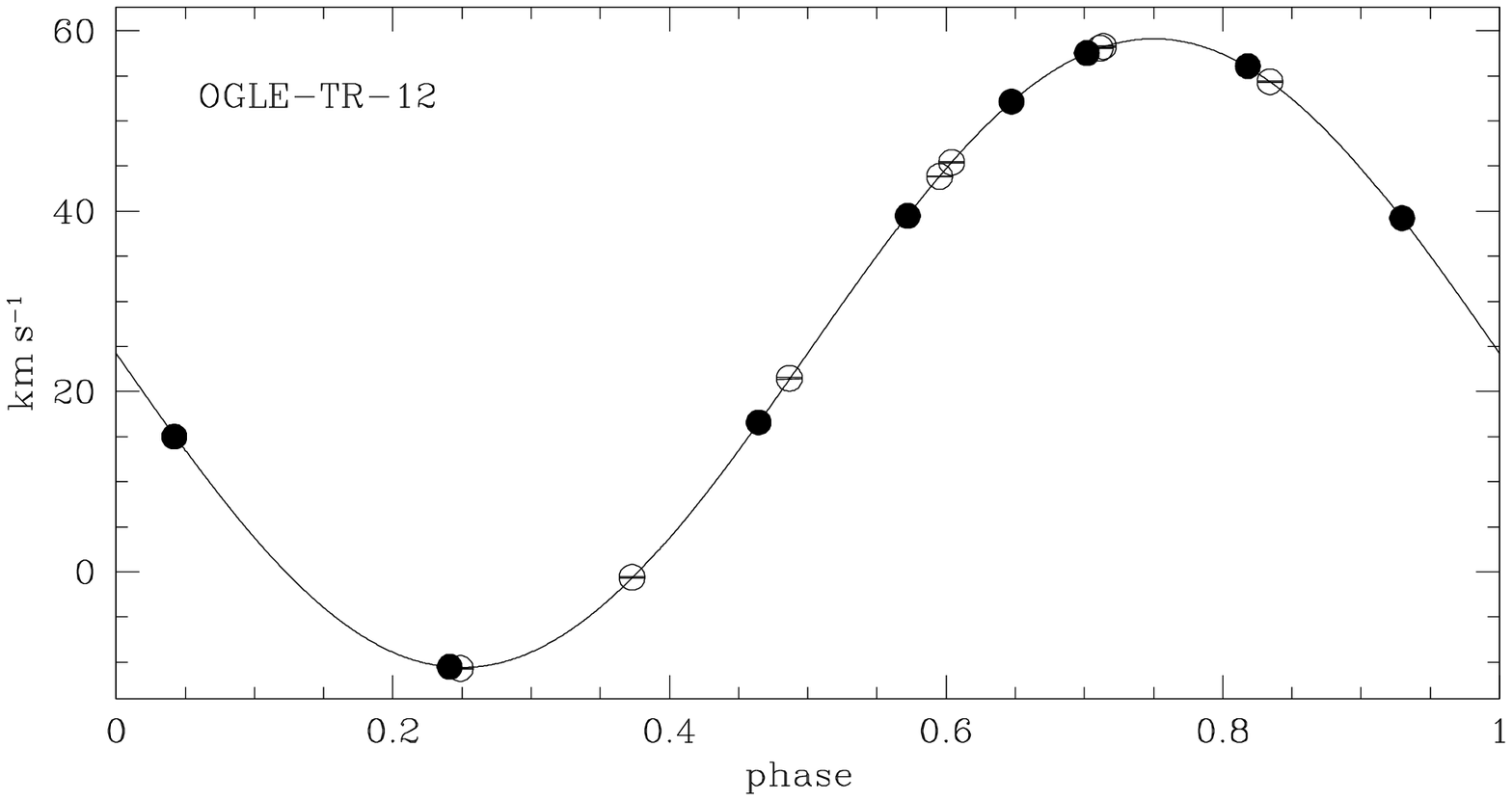}}

\resizebox{8.5cm}{!}{\includegraphics{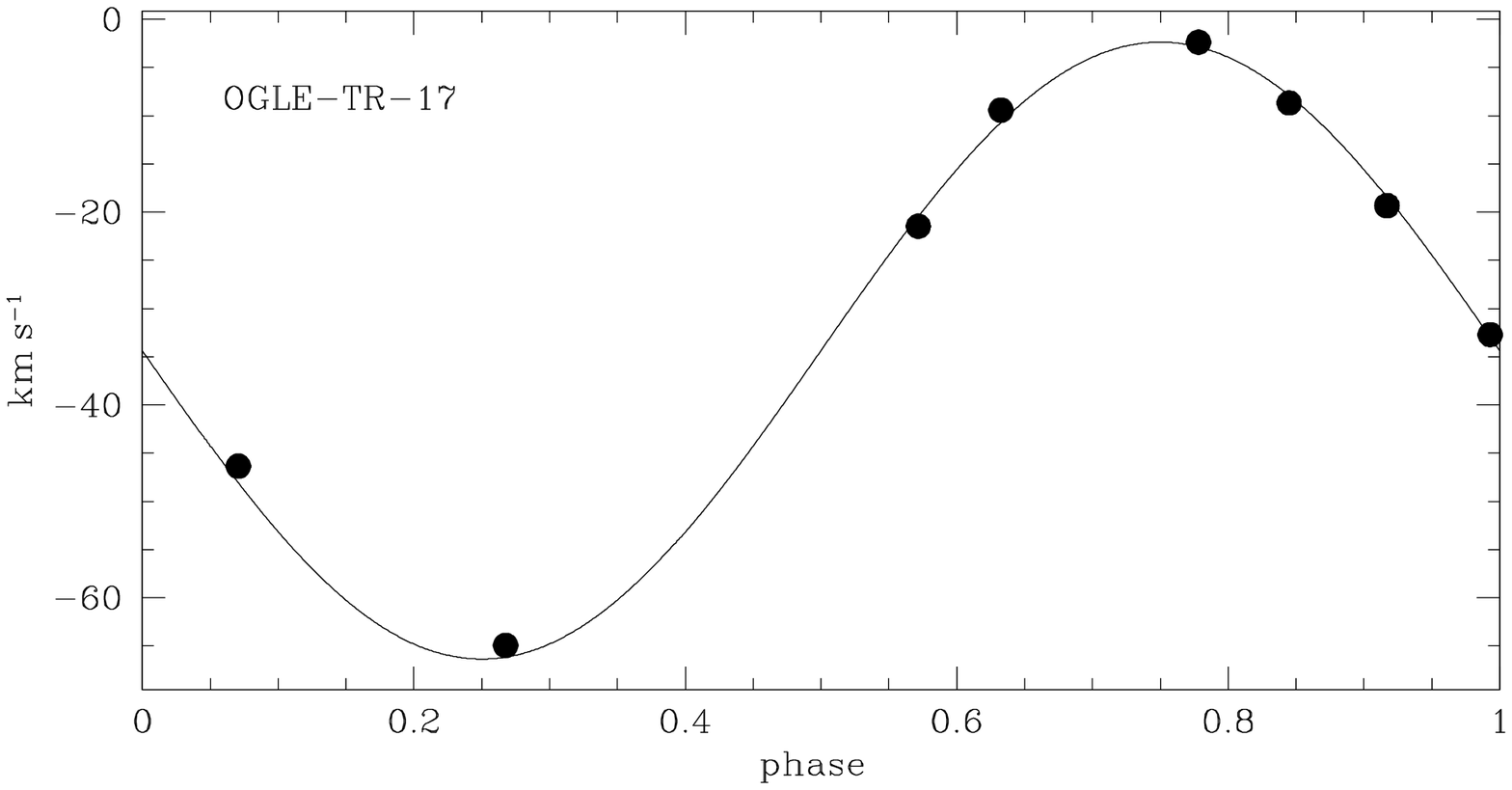}}
\resizebox{8.5cm}{!}{\includegraphics{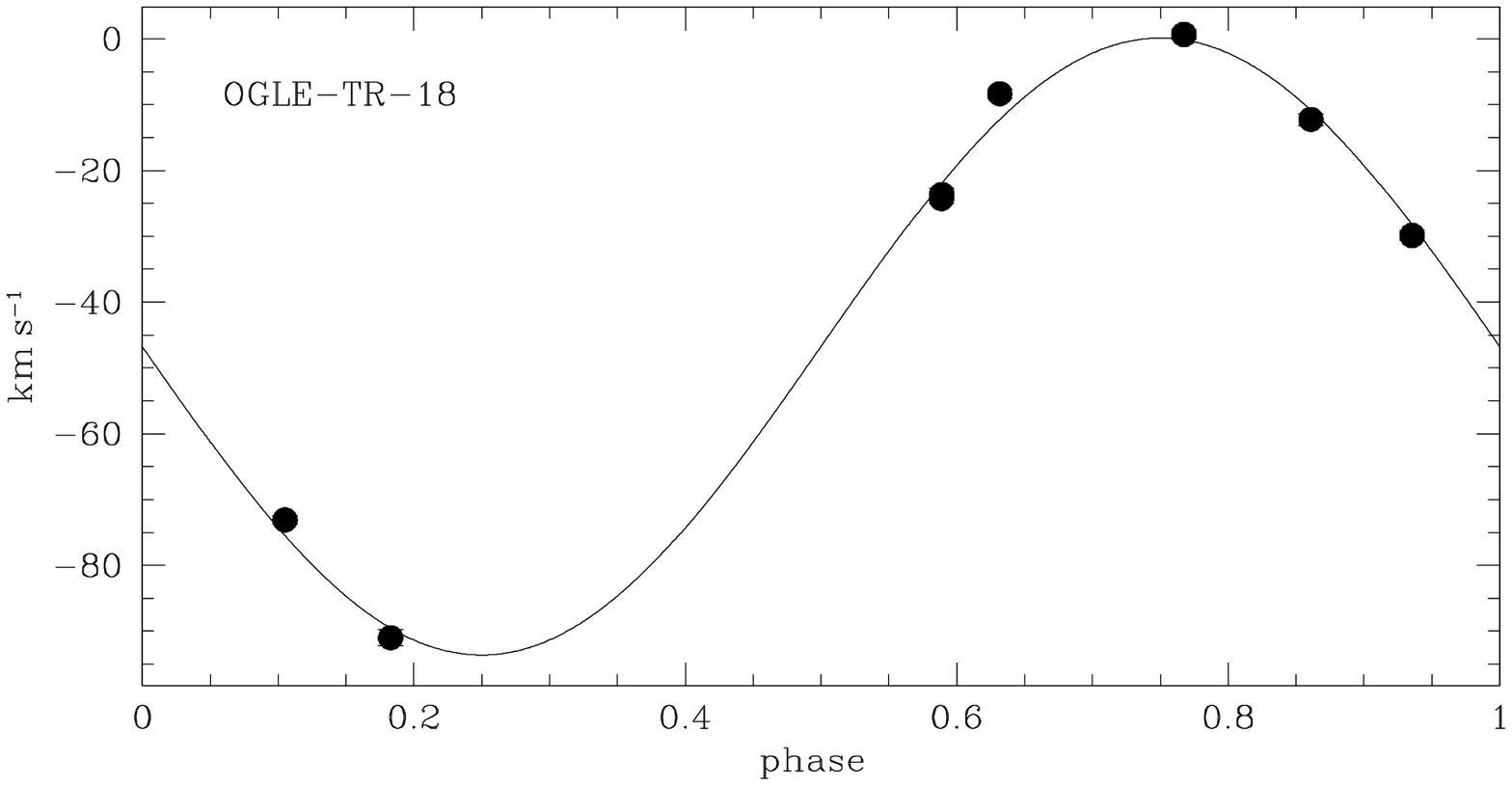}}

\resizebox{8.5cm}{!}{\includegraphics{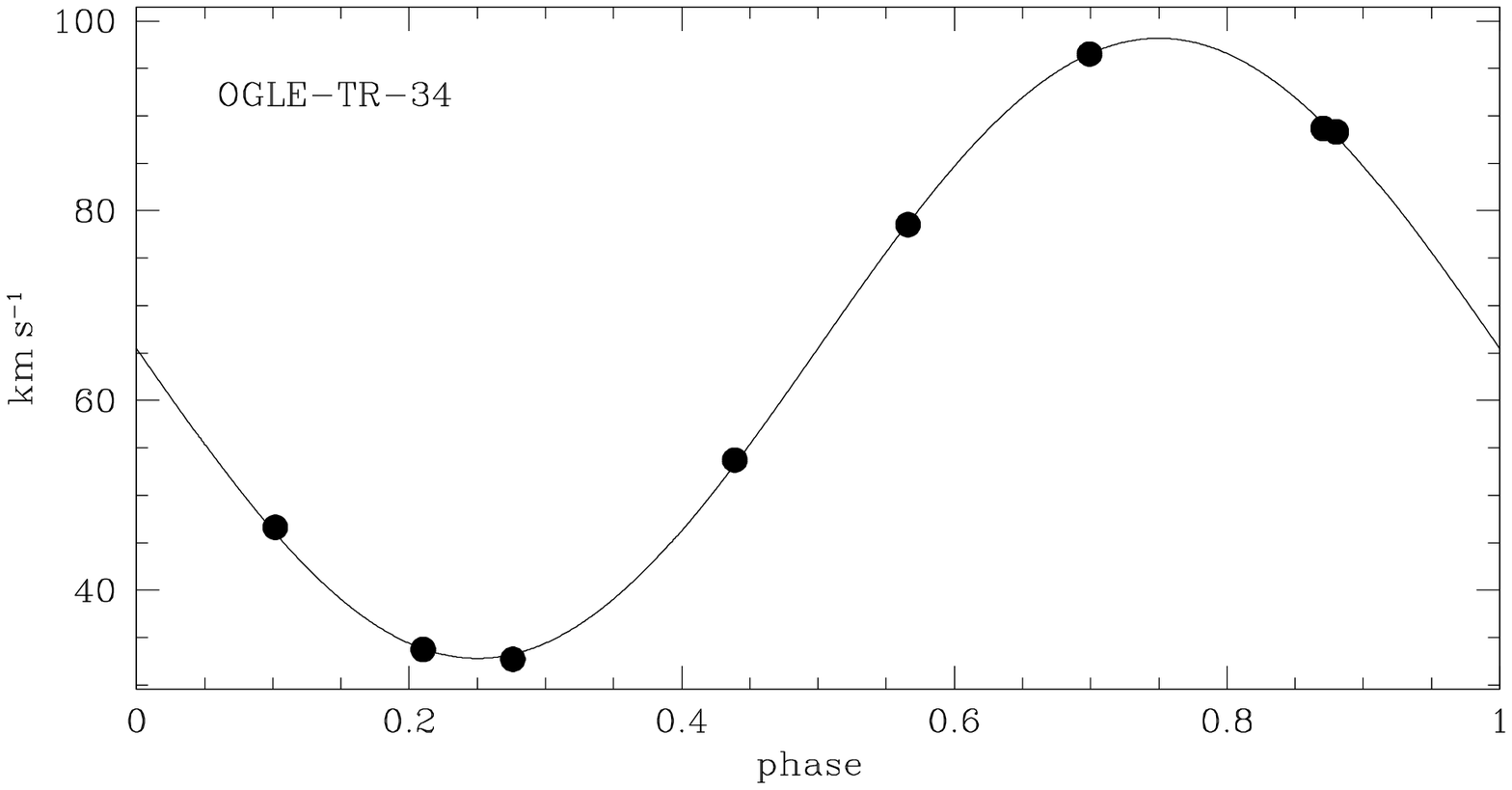}}
\resizebox{8.5cm}{!}{\includegraphics{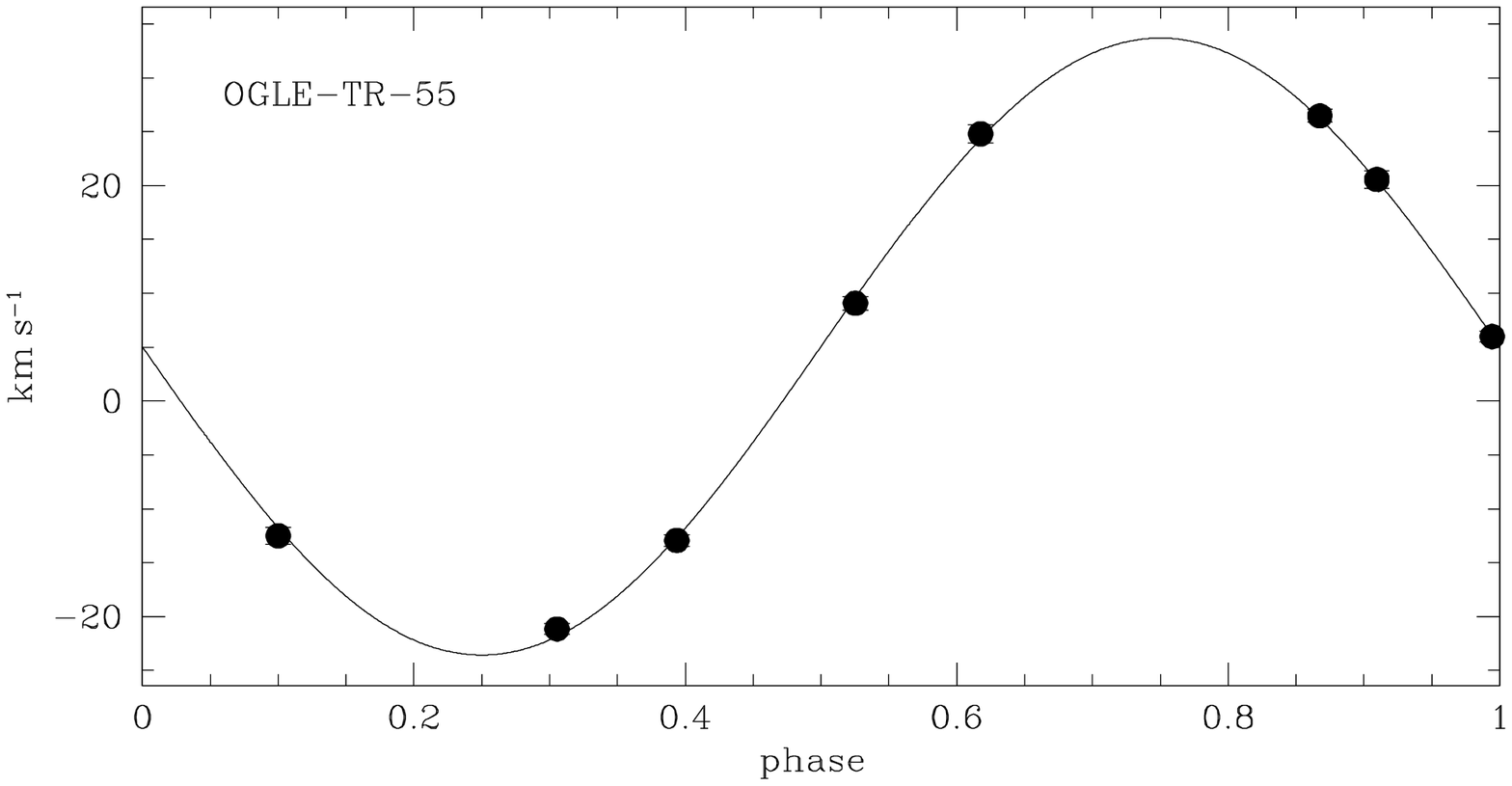}}

\caption{Phase-folded radial velocities of the low mass star transiting companions. 
For OGLE-TR-12, black and white points correspond respectively to FLAMES and UVES measurements.}
\label{sb1}
\end{figure*}

The orbital parameters we derived for the 8 low mass transiting stellar companions OGLE-TR-5, 6, 7, 
12, 17, 18, 34 and 55 are reported in Table~\ref{tableKsb1} and Fig~\ref{sb1}.

\begin{description}

\item[OGLE-TR-5: ]
This candidate is rotating very rapidly ({\vsini}$\sim88$~{\kms}), in synchronization with 
its very short orbital period (0.8 days). Systematic ellipsoidal variations in the light 
curve reveal the deformation of the primary and therefore the massive nature of the 
secondary. The mass obtained from the transit duration and radial velocity semi-amplitude 
is compatible with the spectral type and radius for a G+M binary, and all parameters can be 
determined precisely.

\item[OGLE-TR-6: ] 
The synchronized rotation of this target allows a precise determination of its radius. The 
transit shape shows that the impact parameter is small, and therefore all parameters 
can be computed precisely with our "standard" procedure.

\item[OGLE-TR-7: ]
The synchronized rotation of this target allows a precise determination of its radius. 
The transit shape shows that the impact parameter is small, and therefore all parameters 
can be computed precisely with our "standard" procedure. The computed values of $R$ 
and $M$ are compatible with the spectral type for a main-sequence F dwarf.

\item[OGLE-TR-12: ]
The radial velocity data show that this candidate is a binary system with a period 
1.5 times greater than the period reported by Udalski et al. (\cite{udalski1}). 
This value is fully compatible with the light curve. The rotation is 
synchronized with the orbital period. The light curve indicate that the impact 
parameter $b$ is probably high, causing increased uncertainties on $r$ due to the 
degeneracy between impact parameters and the $\overline{r}$ factor. The large uncertainties 
are of course also due to the fact that there are only two nights in the transit.
OGLE-TR-12 was observed with UVES (white points) and FLAMES (dark points). 
The velocity offset determined between the two set of data indicates a difference 
of 0.200 {\kms} between the two instruments. 

\item[OGLE-TR-17: ]\label{ogle17}
The radial velocity data show that this target is a binary system with P=13.8736 days, 
6 times the value in Udalski et al. (\cite{udalski1}). This revision is compatible 
with the light curve data. The estimated rotation velocity of OGLE-TR-17 seems to 
indicate that this star is not synchronized which is not so surprising considering its 
quite large period. The shape of the transit allows high values of $b$, therefore 
causing degeneracy between the parameters. Moreover, the poor phase coverage of the 
transit increases the uncertainties.
The large O-C residuals clearly indicate a departure from the circular orbit and a 
complete Keplerian fit indicate a small but significant eccentricity of 0.074$\pm$0.008.
Note that OGLE-TR-17, with the longest period, is the only candidate which present a 
significant eccentricity. 

\item[OGLE-TR-18: ]
The rotation velocity is compatible with synchronous rotation. The fit 
of the transit parameters yield a high value of $b$, which increases the uncertainty 
on $r$.  

\item[OGLE-TR-34: ]
The radius given by the rotation velocity assuming synchronization ($R=1.02 \pm 0.04\;R_\odot$) is 
in conflict with the spectral type ($R>1.25\;R_\odot$) and the light curve fit ($R\sim 1.9\;R_\odot$). 
Releasing the synchronism assumption appears to be the most likely possibility 
taking into account the quite large period (8.6 days). In that 
case all the parameters are coherent and well determined. The radius and temperature 
obtained would put OGLE-TR-34 at the evolved edge of the F main sequence. It is possible 
that the radius increase at the end of the main-sequence stage has been too rapid to allow 
synchronization to keep pace, so that the conservation of angular momentum has slowed the 
rotation down to the observed value. Alternatively, the star could be too young for 
synchronization to have taken place yet given the 8.6 days period.

\item[OGLE-TR-55: ]
The rotation velocity shows synchronization with the orbital period. The 
light curve, however, does not constrain the parameters with enough accuracy to confine 
the value of the impact parameter $b$. Because of the large line broadening caused by the 
high rotation velocity, our spectral type analysis yields only a broad estimate of the 
spectral type: F5-G8. The uncertainties on the resulting values $r$ and $m$ are 
correspondingly large. 

\end{description}

\subsection{grazing eclipsing binaries}

\begin{figure*}
\resizebox{8.5cm}{!}{\includegraphics{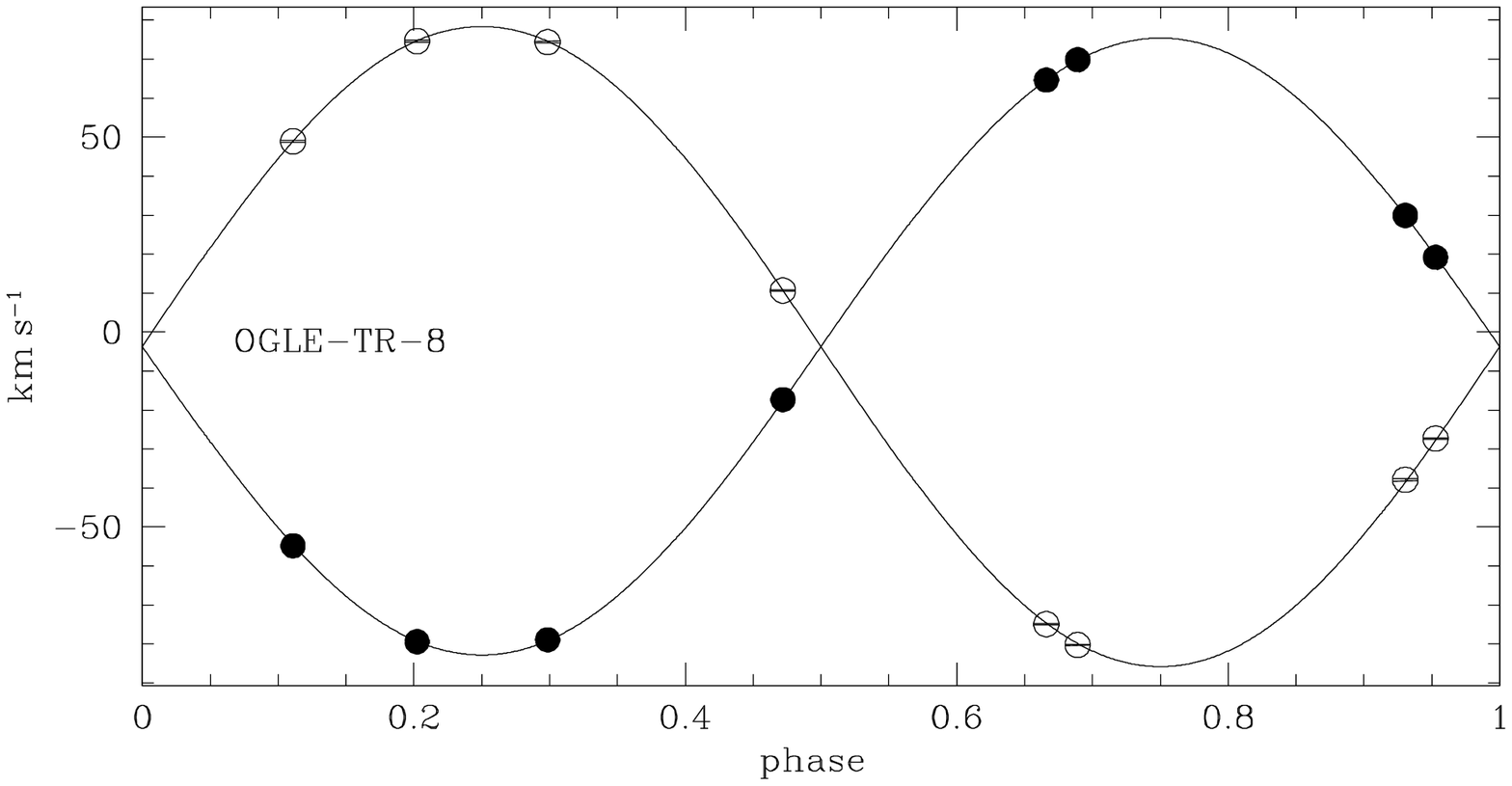}}
\resizebox{8.5cm}{!}{\includegraphics{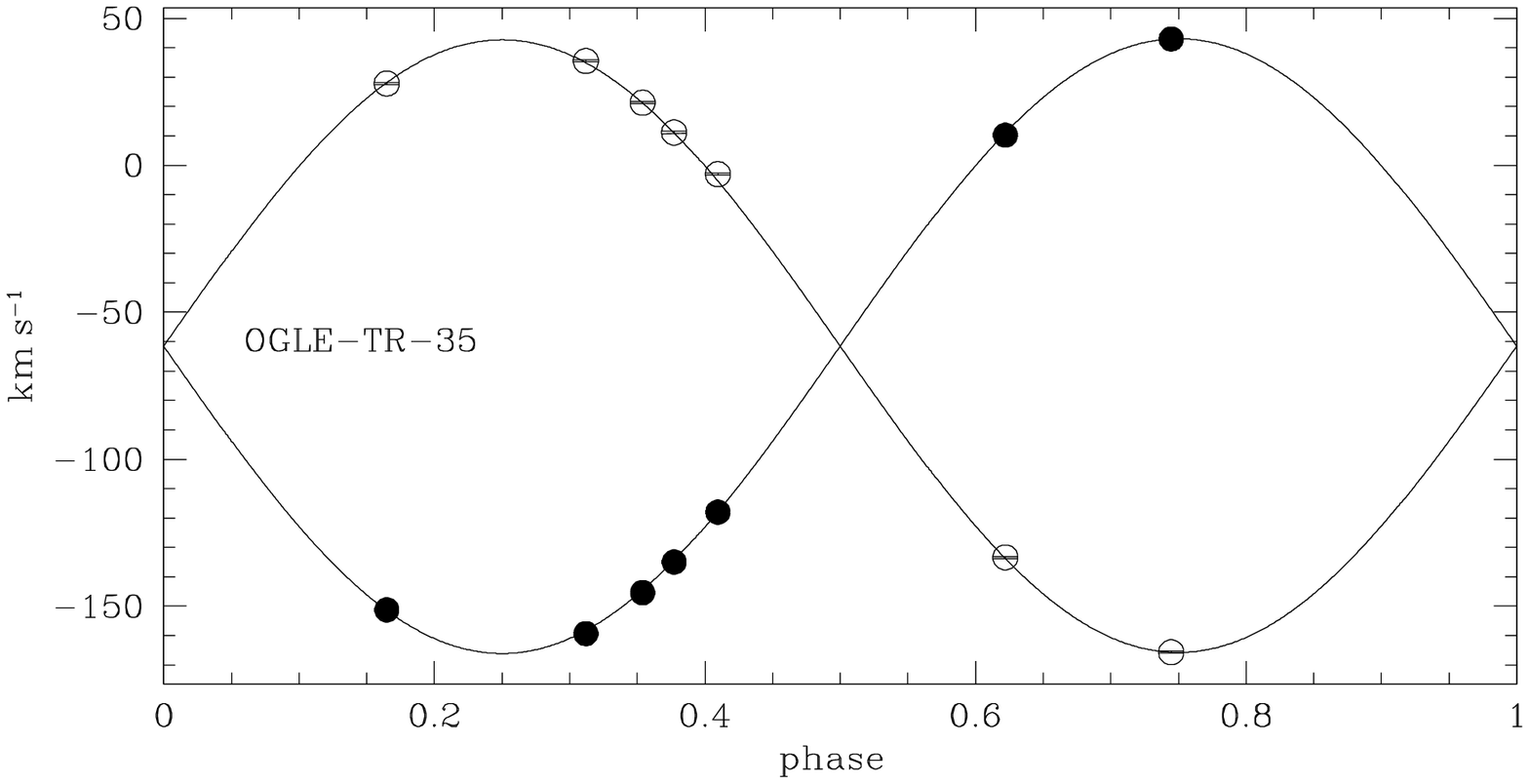}}
\caption{Phase-folded radial velocities of grazing eclipsing binaries. Black and white points 
correspond respectively to component $a$ and $b$.}
\label{sb2}
\end{figure*}

The orbital parameters we derived for the 2 grazing eclipsing binaries OGLE-TR-8 and 35 
are reported in Table~\ref{tableKsb2} and Fig~\ref{sb2}. 

\begin{description}

\item[OGLE-TR-8: ]
The cross-correlation function of the spectra shows two components of 
approximately equal intensity, varying along a SB2 orbit of two times the period given 
by Udalski et al. (\cite{udalski1}), which reveals that both the eclipse and anti-eclipse of 
comparable magnitude were visible in the light curve. The rotation velocities show synchronous 
rotation on both components. The $M \sin i$ of both components are computed from the radial 
velocity curve, the $R \sin i$ from the rotation velocity and period, and the $i$ angle from the 
light curve. The eclipse has to be grazing because the components are of similar size 
and the depth of the eclipse is only a few percent. This constraint alone fixes $\sin i$ 
within a very small interval. All parameters are therefore determined with very high 
accuracy, showing that OGLE-TR-8 is a G-G grazing binary.

\item[OGLE-TR-35: ]
The spectroscopic data show that this target is a double-line binary with a period double 
that given by Udalski et al. (\cite{udalski1}). Therefore, both the eclipse and anti-eclipse are 
visible in the light curve. As for OGLE-TR-8, the combination of the radial velocity 
orbits, the two rotation velocity and the transit depth fix all the parameters very 
precisely. OGLE-TR-35 is a F-F grazing binary.
\end{description}

\subsection{Low mass star transiting companions in triple systems}

\begin{figure*}
\resizebox{8.5cm}{!}{\includegraphics{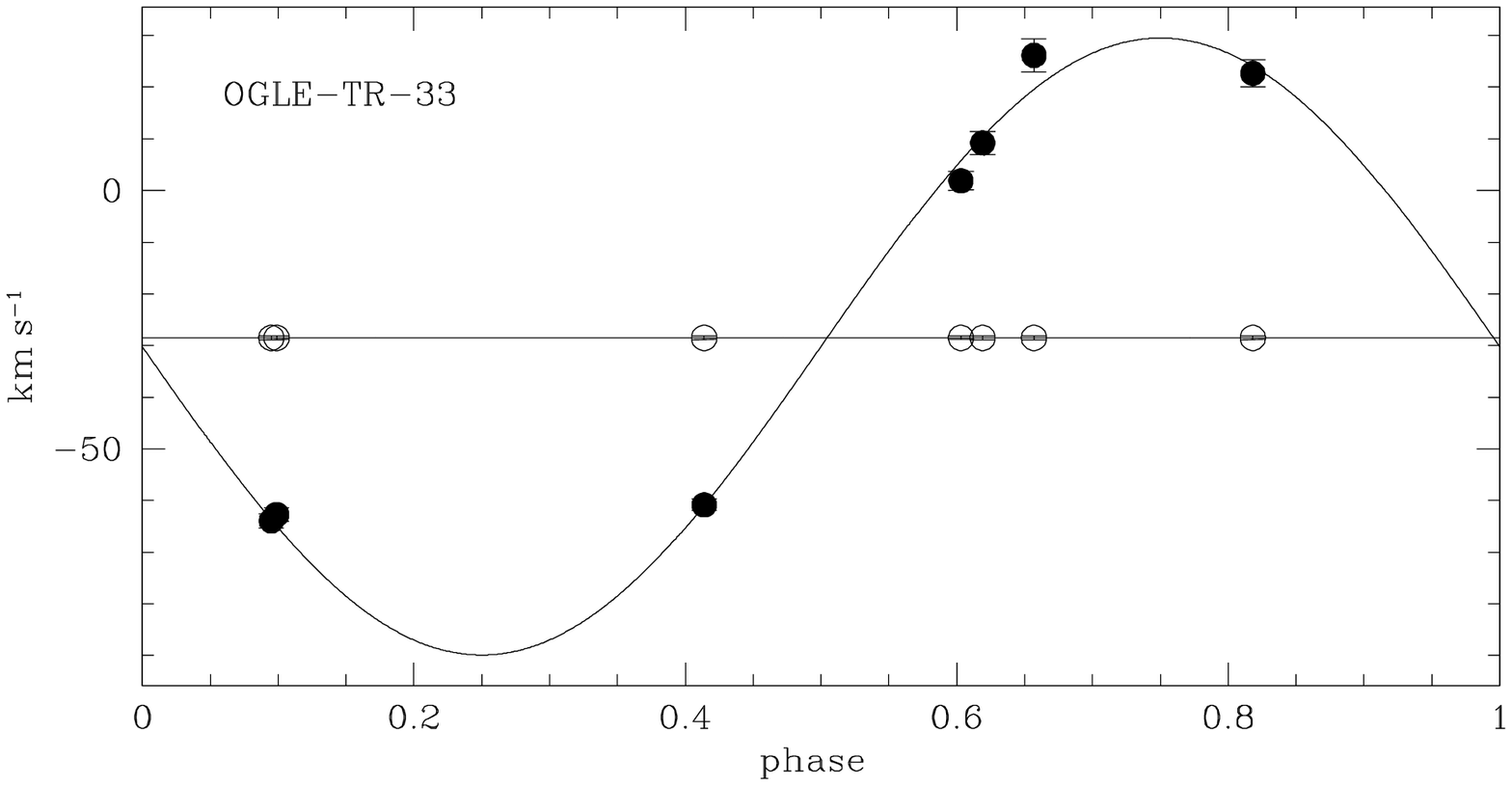}}
\resizebox{8.5cm}{!}{\includegraphics{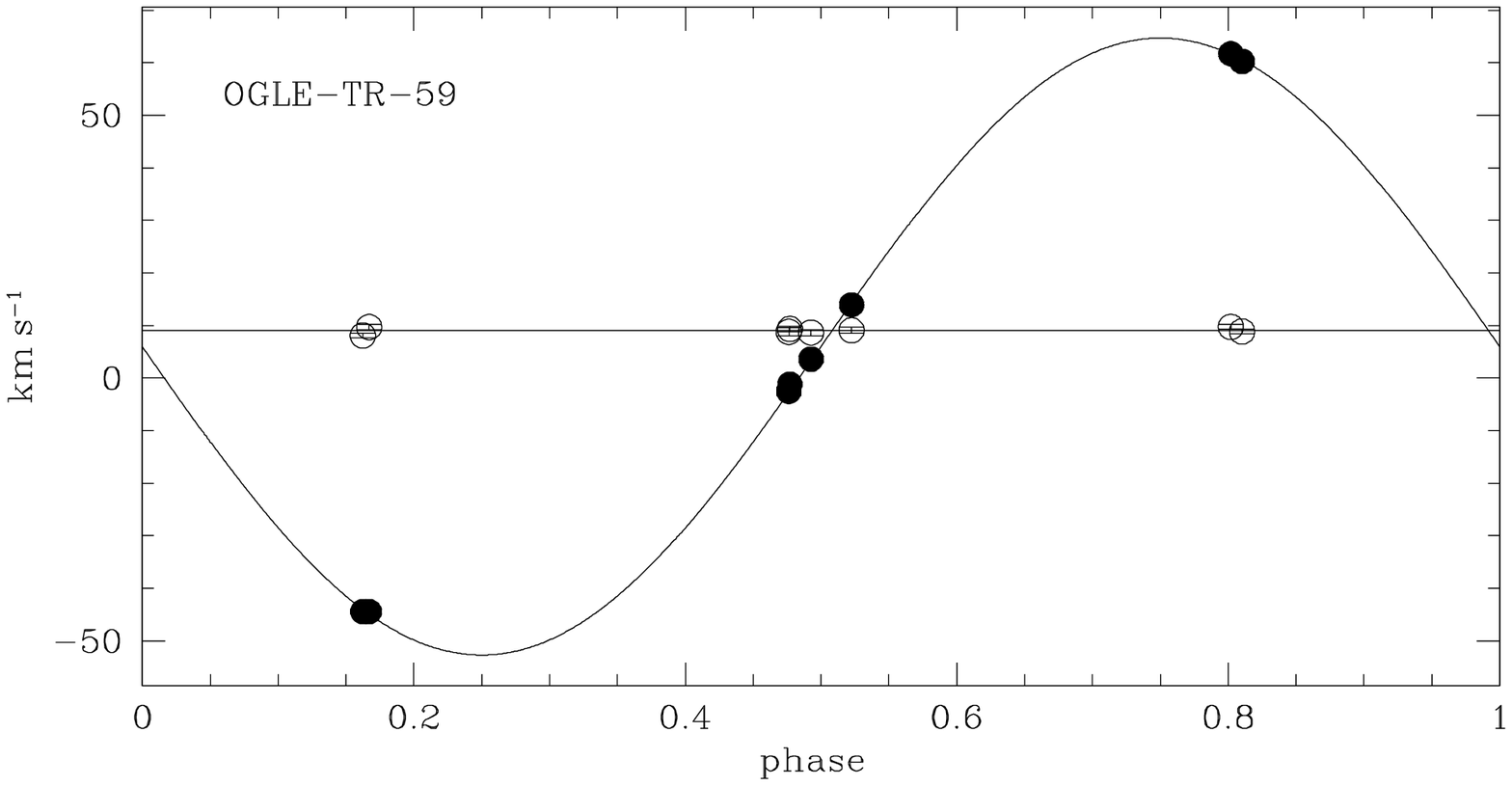}}
\caption{Phase-folded radial velocities of triple system. Black and white points 
correspond respectively to component $b$ and $a$.}
\label{sb3}
\end{figure*}

The orbital parameters we derived for the 2 triple systems OGLE-TR-33 and 59 are reported in 
Table~\ref{tableKsb3} and Fig~\ref{sb3}. To determine the characteristic of the 
second component, we subtracted in the CCF a fixed Gaussian or a fixed rotational profile. 

\begin{description}

\item[OGLE-TR-33: ]
As discussed by Konacki et al. (\cite{konacki2}), this candidate presents a clear blend effect 
visible in the line bisector as well as in the asymmetry of the bottom of the CCF. This 
object consists in a triple system, i.e. an eclipsing binary and a contaminating third 
body. In order to characterize the second spectral component, we subtracted a constant rotational 
profile to the CCF. The second component seems to have a rotation velocity synchronized with 
the orbital period and is in phase with the Udalski et al. (\cite{udalski1}) 
period. The amplitude of the radial velocity variations indicates a low mass star transiting 
companion. A detailed analysis of this system, in full agreement with our result, 
have been made recently by Torres et al. (\cite{torres2}).

\item[OGLE-TR-59: ]
This target has a double-line spectra. However, only one set of lines shows large radial 
velocity variations and presents a period double that given by  Udalski et al. 
(\cite{udalski2}). The transit epoch has to be shifted by half a revised period. Therefore, this 
object consists in a triple system, with an eclipsing binary and a contaminating third 
body, either the component of a physical  triple system, or an unrelated 
background/foreground star. The fixed component represents $\sim70$ percent of the 
CCF surface, and the eclipsed object $30$ percent. If the eclipsed object is in 
synchronous rotation, its rotation velocity implies $R=0.624 \pm 0.033 \;R_\odot$. The transit 
curve is clearly V-shaped, indicating a grazing eclipse, so that the radius ratio of 
the eclipsed component to the eclipsing body is not strongly constrained. If that 
component obeys the $M\sim R$ relation of M dwarfs, then the amplitude of the radial 
velocity variations imply $m\sim0.30\;M_\odot$. Therefore the system would consist of a 
M-M binary with a F/G contaminant in the background. With the present data other 
scenarios are difficult to exclude entirely.

\end{description}

\subsection{Planetary transits, unsolved cases and false positives}
\label{false}

Six of our targets show radial velocity variations lower than 1~{\kms}, or comparable 
with the error bars, indicating the possibility that the transit signal is caused by a 
planet-mass companion. Of these however, only one exhibits the signature of a clear 
orbital motion - the known planetary system OGLE-TR-56 (Konacki et al. \cite{konacki1}, 
Torres et al. \cite{torres}). For OGLE-TR-10, a planetary explanation is proposed. 
For 3 other targets, our data do not allow us to conclude but we strongly suspect a false 
positive transit detection as already examined in section~\ref{existence}. For OGLE-TR-58 
we present strong evidence of a false positive transit detection.

\begin{table*}
\caption{Orbital parameters of planetary candidates, unsolved cases and false positives.}
\label{tableKplanet}
\begin{tabular}{r r r r r r}\hline \hline
Name & P$_{OGLE}$ & T0$_{OGLE}$ & K & V0 & O-C \\ 
 & [days] & $-$2452000 &  [{\kms}] & [{\kms}]& [{\kms}]\\  \hline 
10 & 3.1014 & 70.2190 & 0.081 $\pm$ 0.025 & $-$6.246 $\pm$ 0.017 & 0.068 \\
19 & 5.2821 & 61.89798 & 0.146 $\pm$ 0.048 & $-$33.280 $\pm$ 0.033 & 0.074 \\
48 & 7.2255 & 74.19693 & - & 2.30 $\pm$ 1.45 & 2.9 \\
49 & 2.69042 & 75.53363 & - & $-$106.90 $\pm$ 0.10 & 0.286\\
56 & 1.2119189 & 75.1046 & 0.212 $\pm$ 0.022 & $-$48.324 $\pm$ 0.018 & 0.051 \\
58 & 4.34244 & 76.81982 & - & 51.08 $\pm$ 0.025 & 0.072 \\ \hline
\end{tabular}
\end{table*}

\begin{figure*}
\resizebox{8.5cm}{!}{\includegraphics{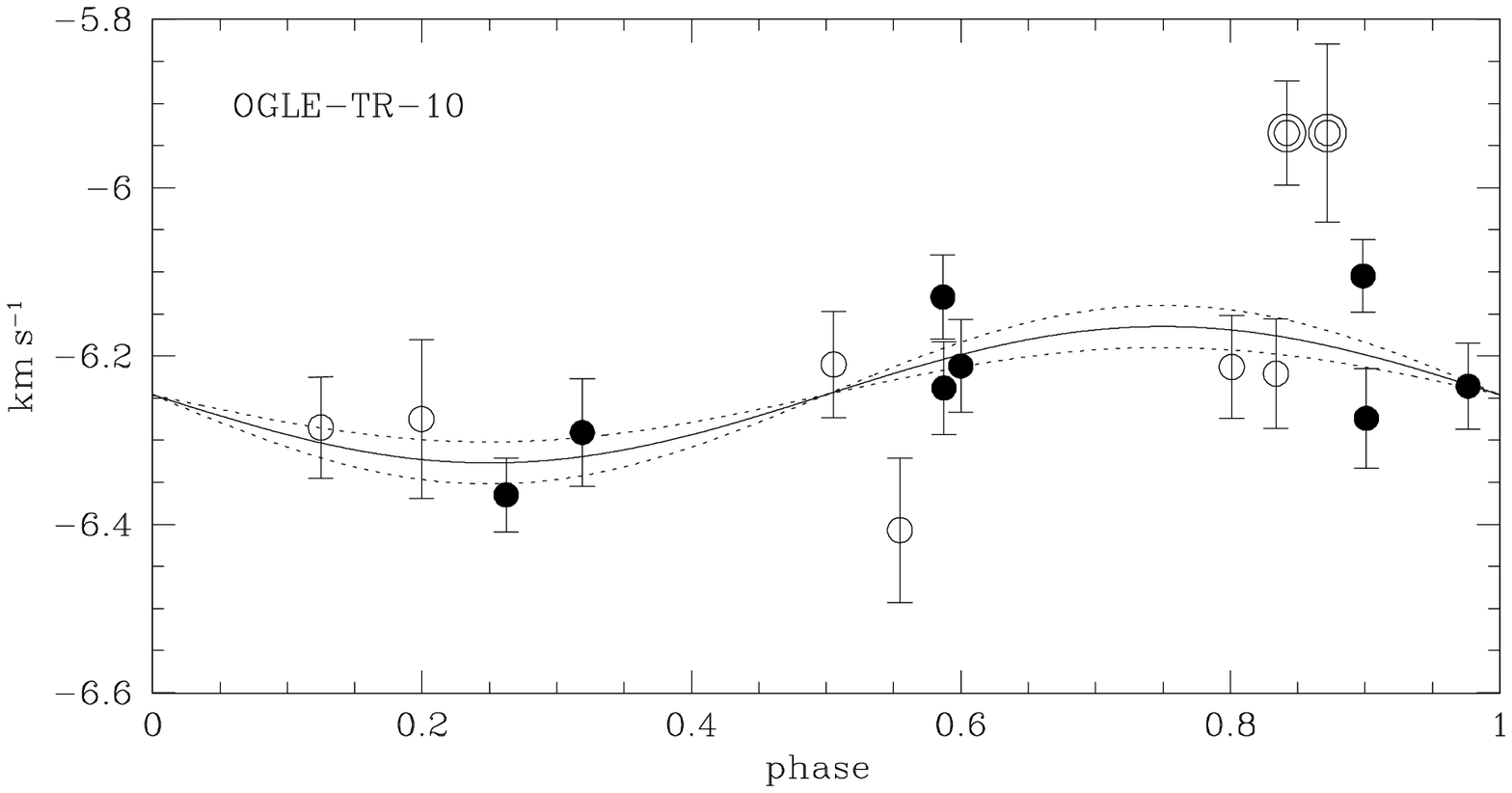}}
\resizebox{8.5cm}{!}{\includegraphics{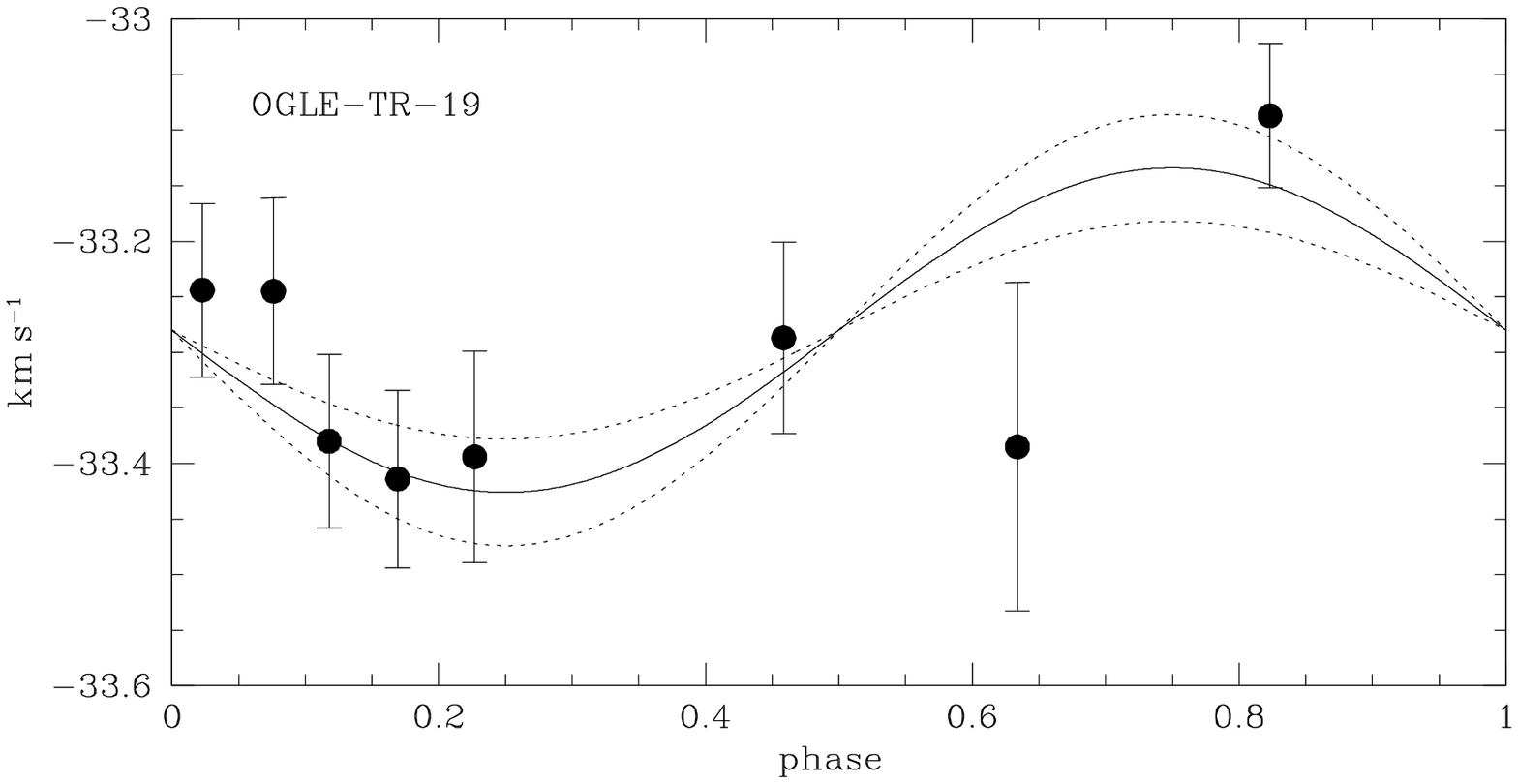}}

\resizebox{8.5cm}{!}{\includegraphics{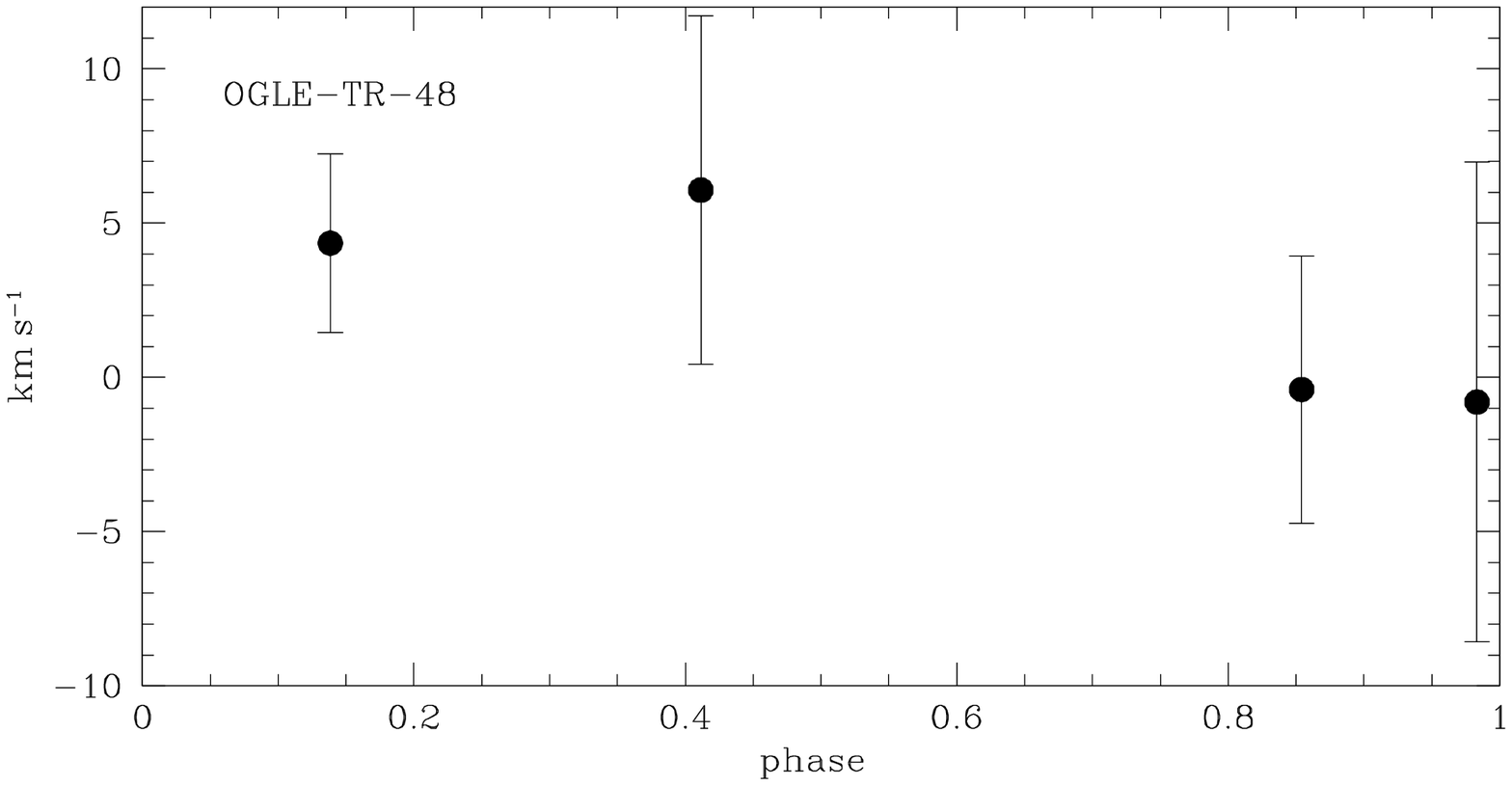}}
\resizebox{8.5cm}{!}{\includegraphics{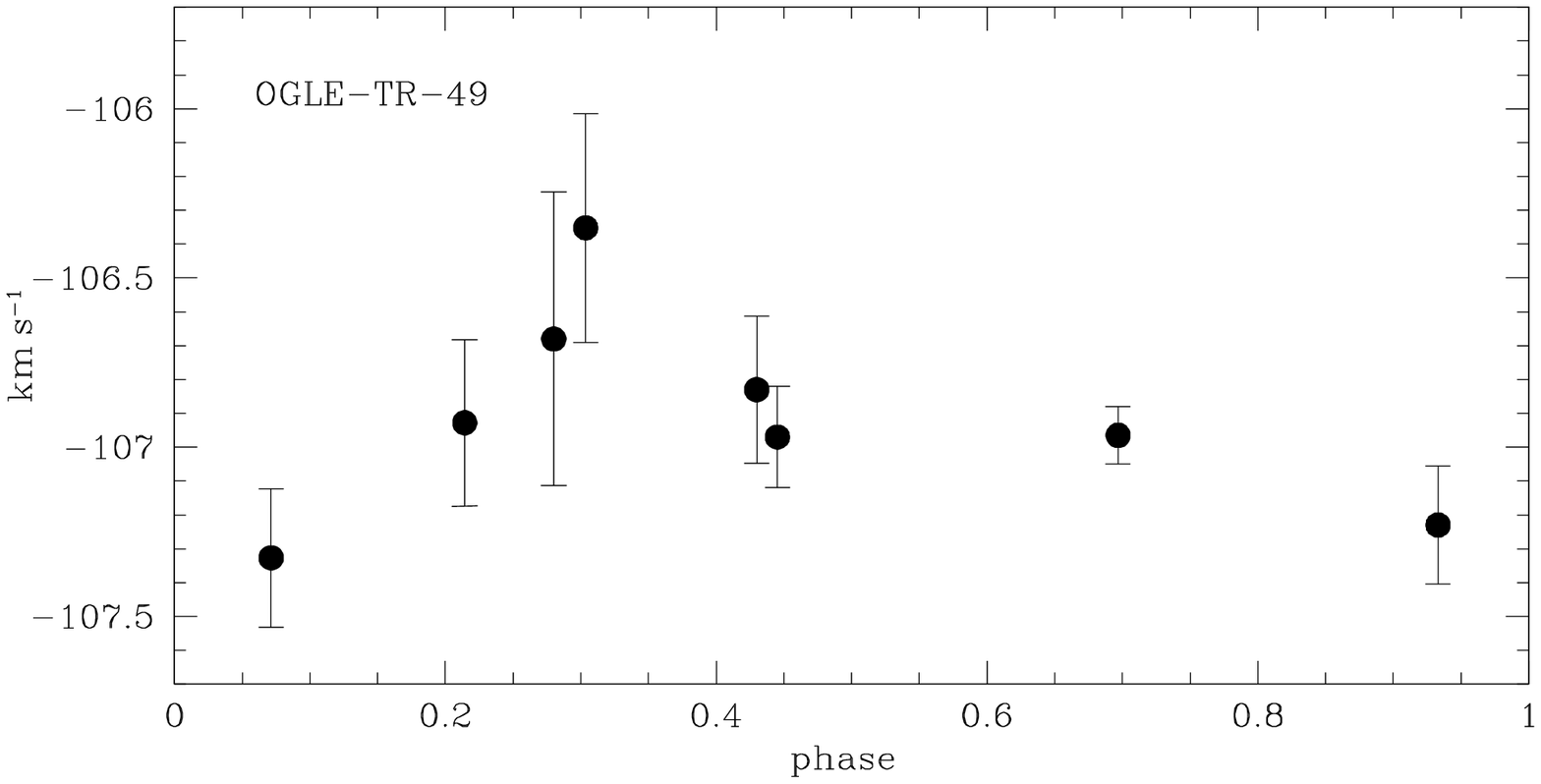}}

\resizebox{8.5cm}{!}{\includegraphics{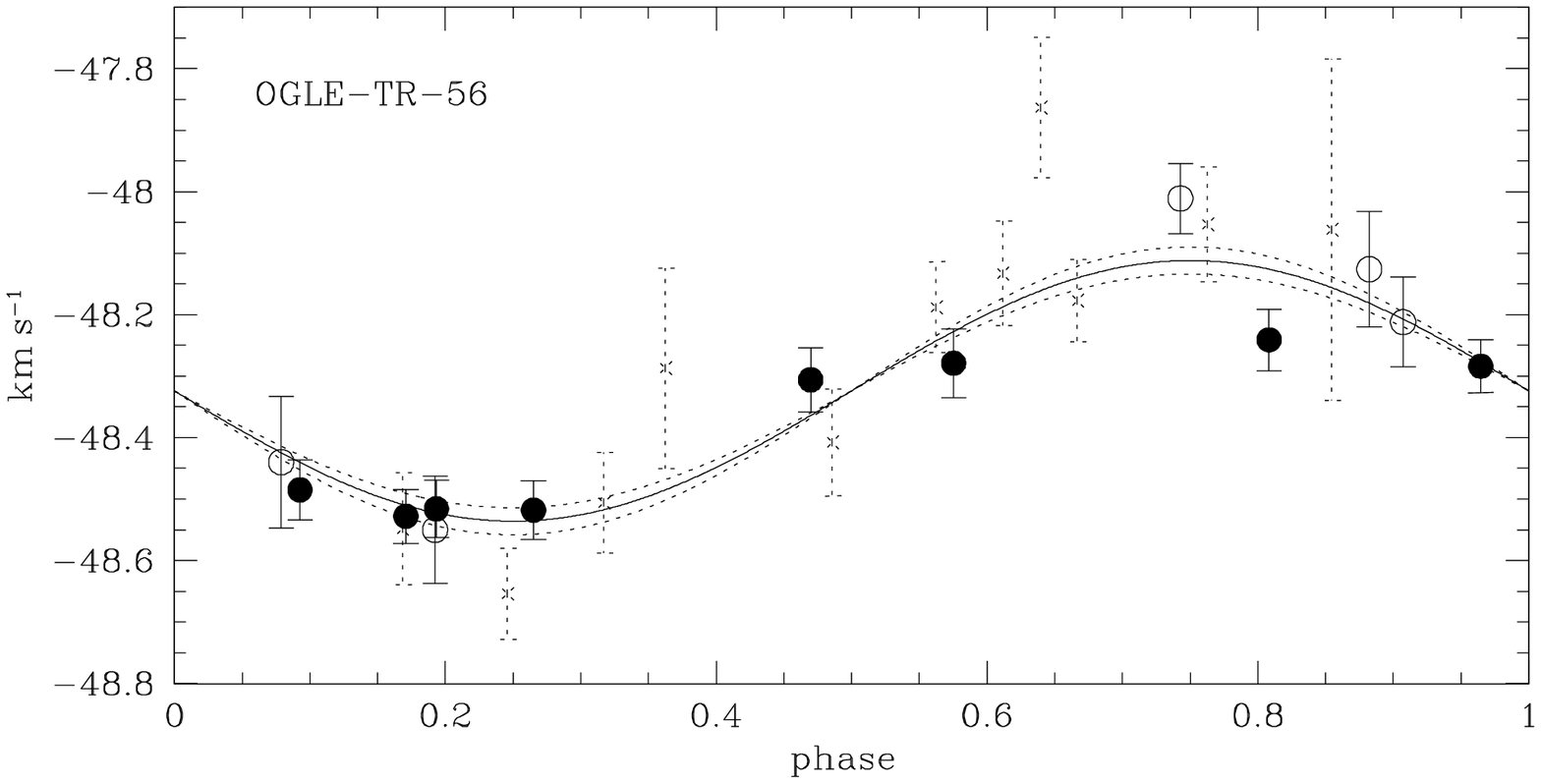}}
\resizebox{8.5cm}{!}{\includegraphics{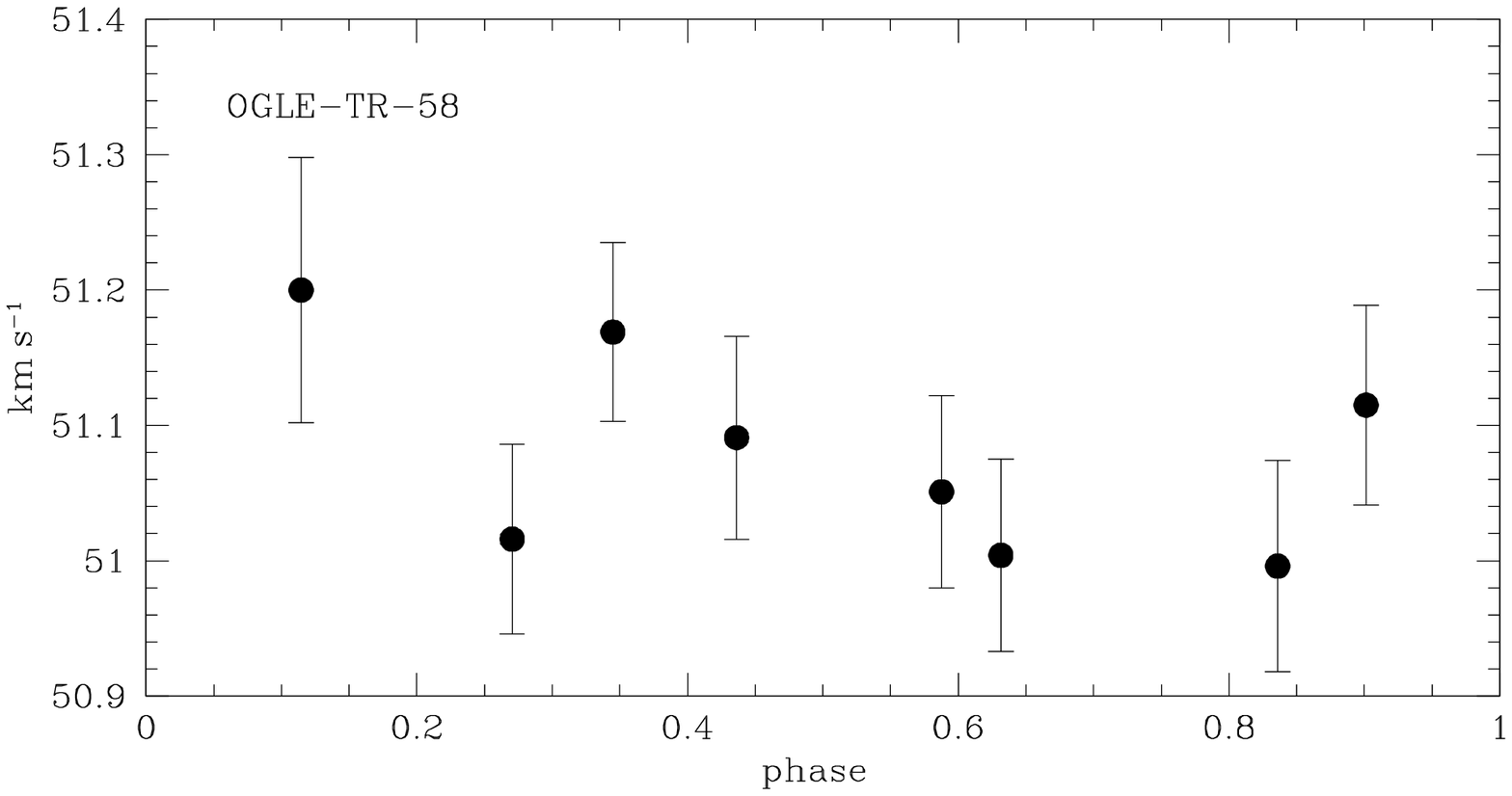}}
\caption{Phase-folded Doppler measurements of planetary candidates and unsolved cases. For OGLE-TR-10, black
and white points correspond respectively to FLAMES and UVES measurements. Encircled points correspond to
measurements made with a seeing lower than 0.9 arcsec. For OGLE-TR-56, black, white and dotted cross points 
correspond respectively to FLAMES, HARPS and Torres et al. (\cite{torres}) measurements. For OGLE-TR-10, 19 and 
56, the dotted lines correspond to fit curves for lower and upper 1-sigma intervals in semi-amplitude K.}
\label{planet}
\end{figure*}

The orbital parameters we derived for the 6 candidates harboring low radial velocity variations 
OGLE-TR-10, 19, 48, 49, 56 and 58 are reported in 
Table~\ref{tableKplanet} and Fig~\ref{planet}. For these candidates we fixed the period given and 
updated by the OGLE team. 

\begin{description}

\item[OGLE-TR-10: ]
This candidate was observed with UVES (white points) and FLAMES (black points). The UVES radial 
velocity was corrected from the offset velocity of 0.200~{\kms} as determined on OGLE-TR-12. Note 
that the two higher values of UVES were made under "unfavorable" seeing conditions (seeing lower 
than 0.9 arcsec) and are not taken into account afterward. The fit indicates an orbital 
signal with $K=81\pm25$ {\ms} which corresponds to a $0.66\pm0.21$ Jupiter mass companion. 
The reduced $\chi^2$ is 2.3 for a constant velocity curve and 1.4 for a circular orbit. The analysis 
of the transit shape and the spectroscopic parameters lead to $r=1.54 \pm 0.12 \; R_{\rm Jup}$.
Such a planet, if confirmed, would have a density quite lower than HD209458. 
It is also quite interesting to note that OGLE-TR-10 presents an excess of metallicity, as 
it is well known that the planet-hosts discovered using radial-velocity techniques are, 
in average, significantly more metal-rich than average field dwarfs (e.g. Santos et al. \cite{santos01}).
In order to examine the possibility that the radial velocity variation is due to a blend scenario, 
we computed the CCF bisectors as described by Santos et al. (\cite{santos02}). We did not 
find any significant bisector effect in the CCF. We also checked the influence of the 
cross-correlation mask used in the CCF computation and did not find any effect. 
However further observations are needed to confirm our hypothesis and a blend scenario 
could not completely be excluded considering the low SNR of our data. 
Both our observations and the possibility of a blend scenario are in agreement with
Konacki et al. (\cite{konacki2}).

\item[OGLE-TR-19: ]
The phase-folded Doppler measurements indicates a significant variation. The reduced $\chi^2$ is 
2.3 for a constant velocity curve and 0.7 for a circular orbit. The fit gives an orbit 
with $K=146\pm48$ {\ms} which correspond to a $1.2\pm0.4$ Jupiter mass companion. The rotation velocity 
is comparable with the instrumental broadening, so that synchronous rotation is excluded. 
The analysis of the transit shape gives $\overline{r}=0.343$ and $b$ in the range [0.79-1.09]. The 
spectroscopic parameters indicate $R=1.46\pm0.16\;R_\odot$ and $M=0.92\pm0.05\;M_\odot$, 
implying $r=4.9 \pm 1.1 \; R_{\rm Jup}$. 
Such a radius is much larger than any planet model would predict, and reinforces 
the odds against the planetary explanation. A single star blended with a background eclipsing binary 
would probably explain the data more convincingly. 
Our bisectors analysis, limited by the high uncertainties, cannot exclude such a scenario. 
Alternatively, we note that only two transits were observed for this candidates 
(see Fig.~\ref{confidence} and Fig.~\ref{doubtful}), which puts OGLE-TR-19 among the less 
secure transit candidates. The "false positive" transit detection could not be excluded 
and only additional photometric and Doppler measurements will allow a definitive conclusion. 
 
\item[OGLE-TR-48: ]
The spectroscopic data reveal that this target is a very rapid rotator, with $V_{rot}>100$ 
{\kms}. Consequently, only very approximate radial velocities could be computed from the 
spectra on only half of the spectra. These measurements do not show significant variations within 
the $\sim 10$ {\kms} accuracy. The rotation is not synchronous with the transit period for any 
reasonable value of the primary radius. We suspect that this very high rotation indicates a 
A or early F star and considering the radius ratio of about 0.14, it seems unlikely that 
the transit could be due to a planetary companion. The upper limit 
on the radial velocity variations exclude a companion with mass larger than $\sim 0.15 \Msol$. 
However, because only two partial transits were observed in the light curve (see Fig.~\ref{confidence} 
and Fig.~\ref{doubtful}), it is not possible to be confident at this stage in the validity of 
the transit period. The most likely explanation is that one or both detected transits are spurious.

\item[OGLE-TR-49: ]
No velocity variation in phase with the transit period and no synchronous rotation 
are observed for this candidate. The reduced $\chi^2$ is 
1.6 for a constant velocity curve.  
Only two transits were observed in the light curve (see Fig.~\ref{confidence}), 
which puts OGLE-TR-49 among the less secure transit candidates. The interval between the two 
transits is $\Delta T=21.52$ days (8 times the period proposed by Udalski et al. 
\cite{udalski2}). If we adopt this largest period, radial velocity gives $K<200$ \ms. 
The transit is poorly defined and the uncertainties on the transit parameters are large. 
The spectroscopic parameters do not allow strong constraint on the mass and radius of the 
primary. Our data can unfortunately not exclude a planetary companion nor an explanation 
in terms of a background eclipsing binary. However, inspection of the light curve 
(see Fig.~\ref{doubtful}) favors an explanation in terms "false positive" transit detection. 

\item[OGLE-TR-56: ]
This candidate was observed with FLAMES (black points) and the HARPS spectrograph (white points). 
The phase-folded Doppler measurements indicate a clear variation. 
The reduced $\chi^2$ is 9.1 for a constant velocity curve and 1.1 for a circular orbit. 
The fit gives an orbit with $K=212 \pm 22 $ {\ms} which correspond to a $1.18\pm0.13$ Jupiter mass 
companion. The analysis of the transit shape and the spectroscopic parameters lead to 
$r=1.25 \pm 0.09 \; R_{\rm Jup}$. We do not find any bisector effect in the CCF nor influence 
of the cross-correlation mask used. Our characterization of OGLE-TR-56b is in agreement (within 
the error bars) with the revised value given by Torres et al. (\cite{torres}) and clearly confirm 
the planetary nature of this object.

\item[OGLE-TR-58: ]
This target shows no radial velocity variation and no synchronous rotation. The reduced 
$\chi^2$ is 1.0 for a constant velocity curve. The suspicion of a spurious transit 
signal in the light curve is even stronger as for 
the previous objects, because the signal detection confidence is the lowest of 
the whole Udalski et al. (\cite{udalski1}, \cite{udalski2}) sample, and the light curve shows 
obvious signs of variability to the level of the hundredth of magnitude. As already noted by 
Konacki et al. (\cite{konacki2}) and Sirko \& Paczynski (\cite{sirko}), the 
mean level has increased by 0.02 mag between one season and the next. Only two possible 
transits were detected, one in a night showing a clear {\it increase} of flux before the 
transit -- so that the average flux over the night is not lower than the normal level for 
OGLE-TR-58 -- and the other consisting of a whole night at lower flux, not a very 
constraining transit signal (see Fig.~\ref{doubtful}). We therefore conclude that OGLE-TR-58 is 
a "false positive" of the transit detection algorithm.

\begin{figure}
\resizebox{8.5cm}{!}{\includegraphics{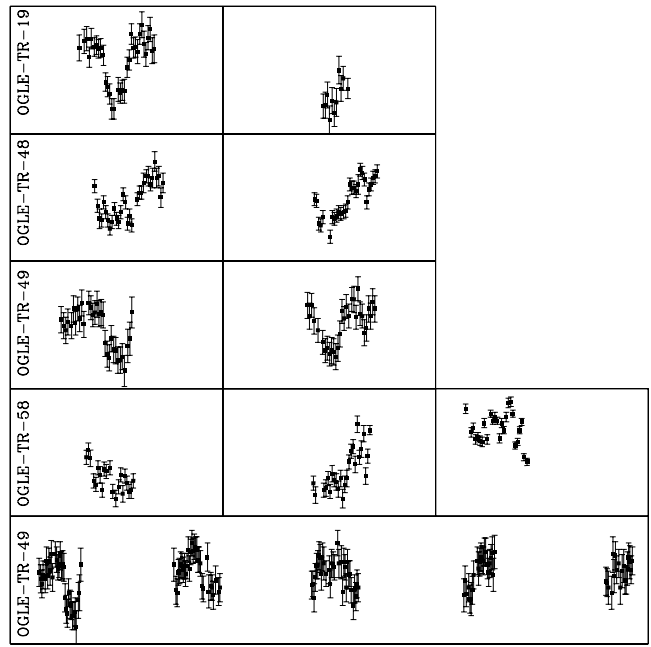}}
\caption{Light curve data during the night of detected transits for OGLE-TR-19, 48, 49, and 58 (from top to bottom) 
and light curve of OGLE-TR-49 for some days after the first detected transit (bottom). The correlation of 
the residuals is clearly visible in the other nights as well.}
\label{doubtful}
\end{figure}

\end{description}

\section{Discussion and conclusion}
\label{discu}

\begin{table*}
\caption{Summary table with the physical parameters $r$, $R$, $m$ and $M$, the orbital 
angle $i$ and the identification of the system.}
\label{tablezoo}
\begin{tabular}{r r r r r r l}\hline \hline 
Name& r & R & m & M & i & comments \\ 
  & [R$_\odot$] & [R$_\odot$] & [M$_\odot$] & [M$_\odot$] & \\ \hline
5 & 0.263 $\pm$ 0.012    & 1.39 $\pm$ 0.026 & 0.271 $\pm$ 0.035 & 1.12 $\pm$ 0.21 & 81-90 & G-M binary \\
6 & 0.393 $\pm$ 0.018    & 1.99 $\pm$ 0.068 & 0.359 $\pm$ 0.025 & 1.37 $\pm$ 0.14 & 88-90 & F-M binary \\
7 & 0.282 $\pm$ 0.013    & 1.64 $\pm$ 0.052 & 0.281 $\pm$ 0.029 & 1.32 $\pm$ 0.21 & 86-90 & F-M binary \\
8 & 1.217 $\pm$ 0.035    & 1.27 $\pm$ 0.035 & 1.160 $\pm$ 0.010 & 1.20 $\pm$ 0.010 & 86 & G-G SB2 \\
10 & 0.156 $\pm$ 0.012   & 1.21 $\pm$ 0.066 & 0.00063 $\pm$ 0.00020 & 1.22 $\pm$ 0.045 & 87-90  & possible planet\\
12 & 0.440 [0.380-1.089] & 1.47 $\pm$ 0.065 & 0.492 $\pm$ 0.026 & 1.28 $\pm$ 0.10 & 85 & F-M binary\\
17 & 0.371 [0.325-0.956] & 1.52 $\pm$ 0.110 & 0.482 $\pm$ 0.040 & 1.06 $\pm$ 0.13 &  86-88 & F-M binary\\
18 & 0.390 $\pm$ 0.040   & 1.95 $\pm$ 0.042 & 0.387 $\pm$ 0.049 & 1.18 $\pm$ 0.22 & 79-86 & F-M binary \\
19 &  -  &  -  &  -  &  -  & -  & unsolved case \\
33 &  -  &  -  &  -  &  -  & -  & blend \\
34 & 0.435 $\pm$ 0.033   & 2.31 $\pm$ 0.174 & 0.509 $\pm$ 0.038 & 1.56 $\pm$ 0.17 & 86-90 & F-M binary \\
35 & 1.74 $\pm$ 0.039    & 1.71 $\pm$ 0.054 & 1.200 $\pm$ 0.009 & 1.19 $\pm$ 0.009 & 80-81 & F-F SB2 \\
48 &  -  &  -  & - &  -  &  -  & unsolved case \\
49 &  -  &  -  & - & -   &  -  & unsolved case \\
55 & 0.209 [0.204-0.989] &  1.92 $\pm$ 0.036   & 0.276 $\pm$ 0.038 & 1.36 $\pm$ 0.28 &  81-90 & F-M binary\\
56 & 0.128 $\pm$ 0.009   & 1.12 $\pm$ 0.069 & 0.00113 $\pm$ 0.00013 & 1.10 $\pm$ 0.078 & 81-83  & confirmed planet \\
58 &  -  & -   &  -  &  -  & -   & false positive \\
59 &   - &  -  &  -  &  -  &  -  & blend \\ \hline
\end{tabular}
\end{table*}

Table~\ref{tablezoo} summarizes all the available information derived for our objects 
following our procedure described in section~\ref{synth}. 
The variety of cases encountered in our sample of eighteen objects is striking and covers
a large part of the bestiary of possible contaminations in the search for planetary 
transits. Target OGLE-TR-5, 6, 7, 12, 17, 18, 34, and 55 have clear resolved orbits 
of eclipsing binaries with a large F/G primary and a small M transiting 
companion. OGLE-TR-8 and 35 have clear resolved orbits of equal-mass, 
grazing eclipsing binaries. OGLE-TR-33 and 59 have resolved orbits of eclipsing binaries 
in a hierarchical triple system. OGLE-TR-56 shows small radial velocity variations in 
agreement with Torres et al. (\cite{torres}) which confirm the planetary nature of the transiting 
companion. OGLE-TR-10 shows small radial velocity variations which could 
be due to planetary companion. OGLE-TR-19, 48 and 49 are unfortunately not yet solved 
but we strongly suspect false positive transit detections. 
OGLE-TR-58 shows no radial velocity variations and its light curve presents 
clear indication that the detected photometric signal is not a bona fide transit.
Note that in some cases (OGLE-TR-12, 17 and 59), the initial period identified from the 
light curve was not correct and that in many cases our secondary radius are very different 
from the initial value of Udalski et al. (\cite{udalski1}, \cite{udalski2}).

Our study has yielded precise radii and masses for a certain 
number of low mass star companions. The mass-radius relation for these objects is given in the 
Fig.~\ref{MRRwide}. We note that no brown-dwarves were detected in our sample 
in agreement with the so-called brown dwarf desert for the short period companions. 
No stellar companions were detected in the mass domain 0.6-1.0~M$_\odot$ due 
to the fact that we selected in priority the smallest candidates of the OGLE survey. 
OGLE-TR12, 17 and 55 (white points) needs additional photometric measurements in order 
to properly constrain the impact parameter $b$ of the transits.
For the other 5 low mass star transiting companions, the precision on the radius and mass 
determination are respectively on the range 4.5-10\% and 7-13\%. This precision is 
not at the level needed to provide a crucial test of stellar physics 
(e.g., Andersen~\cite{andersen}). However 
the empirical Mass-Radius relation remains poorly constrained due to the lack of 
observations of M-type eclipsing binaries. These 5 new candidates significantly increase 
the number of known M-type eclipsing binaries and give new observational constraints on 
models. For comparison we added on Fig.~\ref{MRRwide} the three known M-type 
eclisping binaries (Metcalfe et al.~\cite{metcalfe}, Torres \& Ribas~\cite{torres02}, 
Ribas~\cite{ribas}). Although characterized with a significatively lower accuracy, 
our 5 low mass stars seems to follow the same departure from models.  

The OGLE fields are very crowded, and some of the targets are expected to be contaminated 
by background stars (or foreground fainter stars). Extrapolating the density of bright 
stars to fainter magnitudes indicates that there may be on average 1.4 contaminants stars 
per object down to $\Delta mag =6$. Bright contaminant would be detected by the 
spectroscopy, but faint contaminants can go unnoticed and contribute a few percent to the 
light curve. This makes the photometric transit depth shallower, leading to an 
underestimation of $\overline{r}$. Note that this effect is seeing dependent.

The effect of different assumptions on limb darkening on the derivation of the parameters from 
the light curve were verified using OGLE-TR-6 (central transit) and OGLE-TR-55 (grazing eclipse). 
Changing u1+u2 by 0.2 leads to a difference of the order of 2\% on $\overline{r}$ and on $V_T/R$, while 
removing the limb darkening entirely changes $\overline{r}$ and $V_T/R$ by $\sim$ 8\%.
We also test the effect of changing the limb darkening law in the derivation of the rotation 
velocity from the CCF. Using coefficients  $u1\! + \! u2\! =\! 0.3$ instead of $u1\! + \! u2\! =\! 0.6$ 
modifies $V_{rot} \sin i$ by about 3\%.



\begin{figure*}
\resizebox{15cm}{!}{\includegraphics{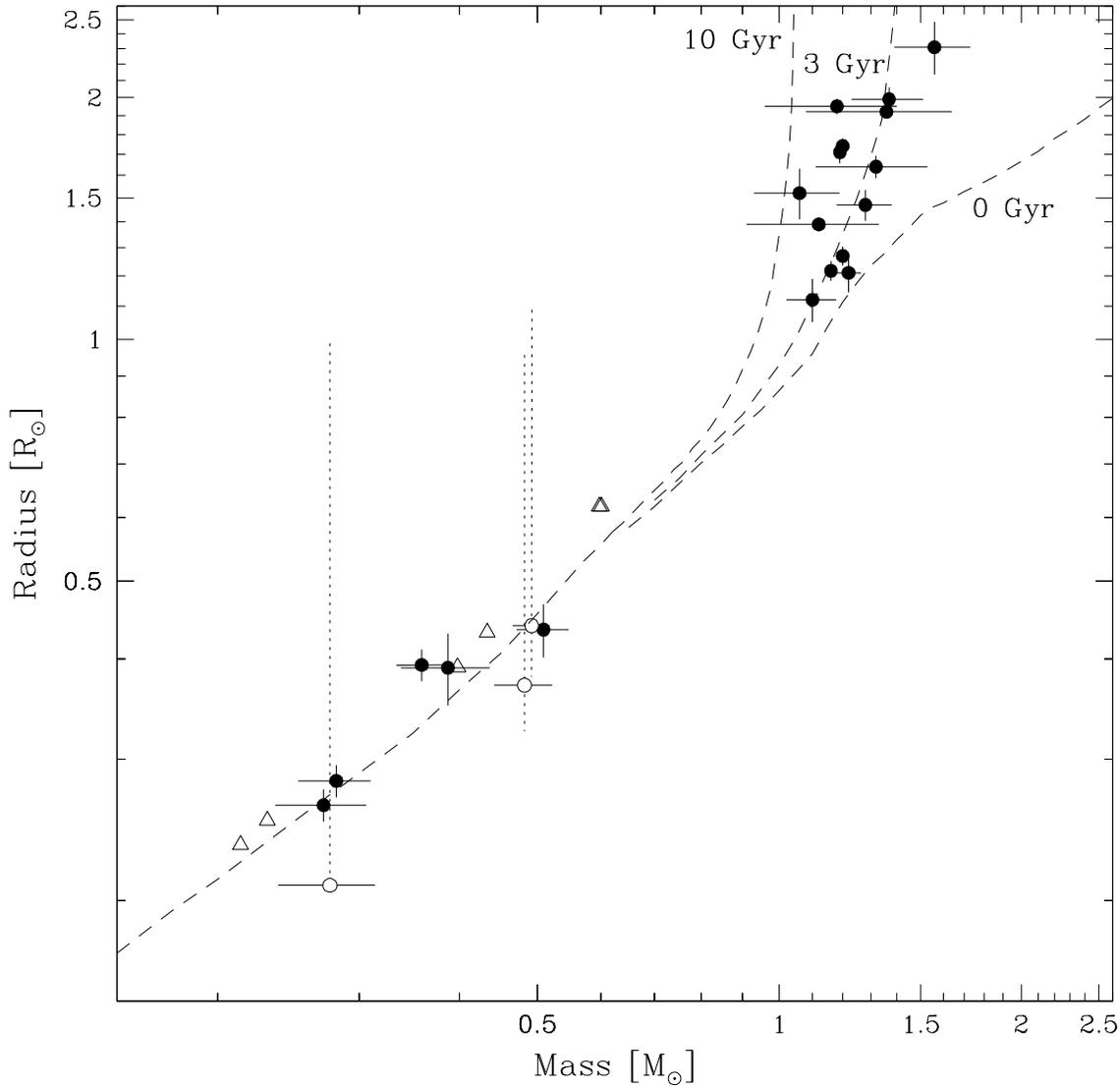}}
\caption{The mass-radius relation for all stellar objects in our sample, primary and 
secondary. Open circles correspond to candidates OGLE-TR-12, 17 and 55 with 
a wide range of possible radius. Triangles correspond to the three known M-type 
eclipsing binaries. The dashed curve up to 0.6 M$_\odot$ corresponds to 
the 1 Gyr theoretical isochrones of Baraffe et al. \cite{baraffe}. Above 0.6 M$_\odot$ the 3 
dashed curves show the Padova model mass-radius relations for solar metallicity 
for three age values.}
\label{MRRwide}
\end{figure*}

This program illustrates the capability of ground-based spectrographs like FLAMES, UVES and HARPS 
to follow the faint transiting candidates found by photometric surveys. It demonstrates the usefulness 
of such a Doppler follow-up in order to discriminate among a large sample of possible 
contaminations in the search for planetary systems. We used in average only 2.5 hours 
of observing time per object, thanks essentially to the very high efficiency of the FLAMES 
multi-fibers facility. Our analysis shows that a large part of the 
transiting candidates could be rejected even a priori through a fine-tuned light curve analysis 
(confidence factor, sinusoidal variations and transit shape). It is also clear that more accurate 
measurements in both photometry and radial velocity will be very useful in order to provide stronger 
constraints on the mass and radius of transiting companions, especially for the suspected 
planetary system OGLE-TR-10, for the unsolved cases OGLE-TR-19 and 49 and for the 
unconstrained radius of the companions of OGLE-TR-12, 17 and 55.

\begin{acknowledgements}
We are grateful to J. Smoker for support on FLAMES at Paranal. F.P. gratefully acknowledges 
the support of CNRS through the fellowship program of CNRS. F.B. acknowledges P. Le Strat for 
continuous support and advices. Support from Funda\c{c}\~ao para a Ci\^encia e Tecnologia 
(Portugal) to N.C.S. in the form of a scholarship is gratefully acknowledged.  
\end{acknowledgements}

\end{document}